\documentclass[%
reprint,
aps,
showkeys,
prd,
nofootinbib,
amsmath,
amssymb
]{revtex4-2}

\usepackage[colorlinks,allcolors=magenta]{hyperref}

\DeclareMathOperator{\Ric}{Ric}
\DeclareMathOperator{\Riem}{Riem}
\DeclareMathOperator{\Pf}{Pf}
\DeclareMathOperator{\Vol}{Vol}

\begin{document}

\title{Universal relations for two- and three-point functions of the CFT energy-momentum tensor and the Weyl anomaly}

\author{Rodrigo Aros}
\email{raros@unab.cl}
\affiliation{Departamento de Fisica y Astronomia, Universidad Andres Bello,\\Av. Republica 252, Santiago, Chile}

\author{Fabrizzio Bugini}
\email{fabrizzio.bugini@unab.cl}
\affiliation{Departamento de Fisica y Astronomia, Universidad Andres Bello,\\Autopista Concepcion-Talcahuano 7100, Talcahuano, Chile}

\author{Danilo E. Diaz}
\email{ddiazvazquez@valenciacollege.edu}
\affiliation{East Campus Science Department, Valencia College,\\701 N Econlockhatchee Trail, Orlando, FL 32825, USA}

\author{Camilo N\'u\~nez-Barra}
\email{cnb@uc.cl}
\affiliation{Facultad de F\'isica, Pontificia Universidad Cat\'olica de Chile,\\Av. Vicu\~na Mackenna 4860, Santiago, Chile}

\date{\today}


\begin{abstract}
We establish universal relations, valid for a generic conformal field theory in even dimension $d\geqslant6$, between the coefficients $\mathcal{A}$, $\mathcal{B}$, $\mathcal{C}$ of the energy-momentum tensor three-point function and the coefficients of the quadratic and cubic terms in the Weyl tensor in the Weyl anomaly. Alternatively, the three-point function coefficients can be replaced with the two-point function coefficient $C_T$ and the two energy flux parameters $t_2$ and $t_4$. Our derivation combines well-known holographic results for $C_T$ with more recent holographic results for $t_2$ and $t_4$, along with a newly obtained Chern--Gauss--Bonnet formula for compact Einstein manifolds. This theorem isolates the Weyl-squared and Weyl-cubed contributions in the relation between the Euler density and the $Q$-curvature, allowing us to identify the relevant quadratic and cubic term unambiguously. We also provide two genuine CFT derivations for $C_T$ based on the renormalization-group running of the $TT$-correlator with respect to the arbitrary but necessary mass scale $\mu$. We revisit several examples to illustrate and validate the general result. These include free scalars, GJMS operators, and holographic duals of higher-derivative gravities. 
\end{abstract}

\maketitle

\section{Introduction}

Conformal field theories (CFTs) occupy a central position in modern theoretical physics. They describe quantum field theories that are invariant under the full conformal group, which extends Poincar\'e symmetry by including scale transformations and special conformal transformations. This enlarged symmetry strongly constrains the structure of correlation functions, the operator spectrum, and the theory's renormalization-group behavior. In any dimension $d$, the local operators of a CFT organize into irreducible representations of the conformal group, labeled by their scaling dimensions and Lorentz spins. Among these, the energy-momentum tensor $T_{ab}$ plays a distinguished role: it generates spacetime symmetries and its correlation functions encode universal dynamical data. In even spacetime dimensions, CFTs exhibit a trace (or Weyl or conformal) anomaly when coupled to a background metric~\cite{Capper:1974ic}. Although the classical theory is Weyl invariant, the quantum effective action acquires a non-vanishing trace proportional to local curvature invariants. 

\subsection{Energy-momentum one-point function}

The anomaly takes the schematic form~\cite{Deser:1993yx,Boulanger:2007st,Boulanger:2007ab} 
\begin{equation}
    (4\pi)^{d/2}\langle T\rangle=-aE_d+\sum_ic_iI_i+\text{ttd},
\end{equation} 
where the Euler density $E_d$ carries the type-A central charge $a$, the pointwise Weyl invariants $I_i$ carry the type-B central charges $c_i$, and $\text{ttd}$ stands for trivial anomalies or trivial total derivatives whose conformal primitives are local curvature invariants. The central charges, traditionally denoted $a$ and $c$ in four dimensions and, by suitable generalizations, in higher even dimensions, are intrinsic CFT data.

The trace anomaly anchors the irreversibility of the renormalization group (RG), fixes stress-tensor data in CFTs, generates (non-local) quantum effective actions, controls vacuum, Casimir or Hawking effects in curved space and shapes quantum corrections in cosmology and black hole physics, among many other physical applications in quantum gravity, inflationary cosmology, string theory and statistical mechanics (see, e.g.,~\cite{Duff2017} for a comprehensive review).

\subsection{Energy-momentum two-point function}

In flat space, the two-point function of $T_{ab}$ is fixed up to an overall coefficient $C_T$, which serves as a measure of the number of degrees of freedom of the theory and appears in a variety of physical contexts, from entanglement entropy to conformal collider bounds.  In our conventions, the two-point function of the energy-momentum tensor in flat space reads
\begin{equation}
  S_d^2\,\langle T_{ab}(x) T_{cd}(0)\rangle=C_T\,\frac{\mathcal{I}_{ab,cd}(x)}{|x|^{2d}},\nonumber
\end{equation}
where $S_d=2\pi^{d/2}/\Gamma(d/2)$ is the area of the unit $(d-1)$-sphere, $\mathcal{I}_{ab,cd}(x)$ represents the inversion operator on symmetric traceless tensors $\mathcal{I}_{ab,cd}=(I_{ac}I_{bd}+I_{ad}I_{bc})/2-\delta_{ab}\delta_{cd}/d$, which is made of the inversion tensor $I_{ab}=\delta_{ab}-x_ax_b/|x|^2$, and $|x|$ denotes the Euclidean distance between the two points (see, e.g.,~\cite{Osborn:1993cr}).

The connection between $C_T$ and the Weyl anomaly coefficients is rich and intricate. Although these quantities arise from different physical considerations \emph{flat space correlation functions versus curved space anomalies} they are not independent. In holographic theories, for example, both can be computed from the same bulk gravitational action, leading to universal relations. More generally, following the lead of Osborn and Petkou~\cite{Osborn:1993cr}, since the integrated trace anomaly determines the running of the CFT partition function with the mass scale $\mu$, this ties the Weyl-squared term in the anomaly to the normalization of the stress‑tensor two‑point function around flat space. Thus, in any even $d$, $C_T$ is fixed by the part of the Weyl anomaly that is quadratic in the metric fluctuation, and everything else, such as higher powers of the Weyl tensor, trivial total derivatives, and topological terms, does not feed into $C_T$. Here is how that plays out dimension by dimension.

$\mathbf{d=2}$. There is only one central charge, $c$, and it is defined by
\begin{equation}
    4\pi\langle T\rangle=\frac{c}{6}R.
\end{equation}
This controls both the trace anomaly and the two-point function given by
\begin{equation}
    C_T=2\,c\qquad(d=2).
\end{equation}
This result, of course, complies with the well-known results for the conformal scalar $C_T=d/(d-1)=2$ and the convention that $c=1$ for the free boson.

\textbf{$\mathbf{d=4}$.} The anomaly is determined by $a$ and $c$, coefficients of the Euler density and Weyl tensor squared terms, respectively, modulo a trivial total derivative,
\begin{equation}
    (4\pi)^2\langle T\rangle=-aE_4+cW^2+\text{ttd}.\nonumber
\end{equation}
Here, $C_T$ is uniquely proportional to the $c$ coefficient associated with the square of the Weyl tensor. The two-point function of $T_{ab}$ is controlled by the coefficient of $W^2$, i.e., by $c$. The Euler term $E_4$ and the trivial total derivative do not contribute at quadratic order in the metric fluctuation around flat space, and, as shown in~\cite{Osborn:1993cr},  
\begin{equation}
    C_T=160\,c\qquad (d=4).
\end{equation}

\textbf{$\mathbf{d=6}$.} As $d$ increases, the number of independent Weyl invariants ($I_i$) grows. In $d=6$, there are three such coefficients ($c_1$, $c_2$ and $c_3$). Out of the three Weyl invariants, only one is related to $C_T$. The anomaly is usually written as  
\begin{equation}
(4\pi)^3\langle T\rangle=-aE_6+c_1I_1+c_2I_2+c_3I_3+\text{ttd},\nonumber
\end{equation}
where $I_1$ and $I_2$ are cubic in the Weyl tensor, while $I_3$ contains terms of the schematic form $W \nabla^2W+\text{lot}+\text{ttd}$. The lower-order terms in derivatives ($\text{lot}$) clearly do not contribute to the quadratic expansion around flat space and are not directly related to $C_T$. Therefore, it is the central charge $c_3$ that relates to $C_T$~\cite{Beccaria:2015ypa},
\begin{equation}
C_T=3024\,c_3\qquad (d=6).
\end{equation}

\textbf{$\mathbf{d=8}$.} The pattern expands even further. There is a larger number of Weyl invariants $I_i$ in the anomaly. Fortunately, only those invariants whose expansion contains quadratic terms of the Weyl tensor (with derivatives allowed) can contribute to $C_T$. Strictly cubic and higher curvature invariants start only at cubic order in the metric fluctuation and thus only affect $3$‑point and higher stress‑tensor correlators.

In holographic Einstein gravity, where the bulk action is just the Einstein--Hilbert action plus a negative cosmological constant, the coefficients of the holographic Weyl anomaly~\cite{Henningson:1998ey,Henningson:1998gx} (see also~\cite{Gover:2002ay,Jia:2021hgy,Boulanger:2025oli}) are all determined by a single bulk coupling: that of the critical $Q$-curvature~\cite{Graham2003}. This enforces linear relations among the various $c_i$ and $a$, but structurally, $C_T$ still tracks the unique quadratic-in-Weyl direction in the space of invariants, just as in 6d it tracks $c_3$.

In 8d, Chen and L\"u~\cite{Chen:2024kuw} (see also~\cite{Boulanger:2004zf,Case:2024oih}) have recently found a quadratic in Weyl invariant $I^{(8)}_{10}$ whose coefficient $c_{10}$ becomes equal to $a$ in holographic Einstein gravity. Comparing with the long-known result of Liu and Tseytlin for $C_T$~\cite{Liu:1998bu}, they obtain (in our conventions) 
\begin{equation}
    C_T=69120\,c_{\scriptscriptstyle10}\qquad (d=8).
\end{equation}

\textbf{$\mathbf{d>2}$, even.} In higher dimensions, the precise structure of the type-B Weyl invariants is largely unknown, but for our present purposes, we can safely focus on the term quadratic in the Weyl tensor. So, up to higher curvature terms and trivial total derivatives, we have   
\begin{equation}
    (4\pi)^{d/2}\langle T\rangle=-aE_d+c_{2,1}\frac{d}{4}W^{(d)}_{2,1}+\text{lot}+\text{ttd},
\end{equation}
where we have chosen a particular point-wise Weyl invariant $W^{(d)}_{2,1}$, recently introduced by Case et al.~\cite{Case:2024oih}, which descends from the ambient $(\tilde{\nabla}^2)^{d/2-2}\tilde{W}^2$ in the Fefferman--Graham program. Restricted to Einstein manifolds, it enters an explicit expression for the Chern--Gauss--Bonnet formula that allows for the interchange between the Euler density and the $Q$-curvature. With this insight, we tune the coefficient $c_{2,1}$ in the anomaly to enforce the equality $c_{2,1}=a$ when the anomaly reduces to only the critical $Q$-curvature.

\noindent Alternatively, we are also free to choose instead a different Weyl invariant quadratic in the Weyl tensor that naturally extends the $W^2$ and the $I_3$ from 4d and 6d, respectively (see, e.g.,~\cite{Bastianelli:2000hi}). The numerical coefficient is obtained by the expansion of the $Q$-curvature to terms quadratic in curvature (see subsection~\ref{sec:WeylSquared}) to enforce the equality $c_{2,1}=a$ for the $Q$-curvature, 
\begin{multline}
(4\pi)^{d/2}\langle T\rangle=-aE_d+c_{2,1}\frac{d(d-2)}{8(d-3)}\bigr[W(\nabla^2)^{d/2-2}W\\+ \text{lot}\bigr]+\text{ttd}.
\end{multline}

In what follows, we explain how to obtain a universal relation between $C_T$ and $c_{2,1}$ that accommodates low dimensions and extrapolates to any even higher dimension. This general result can be modified to include the two-dimensional case by absorbing a factor $6$ in the definition of the 2d $c$, setting instead $4\pi\langle T\rangle=cR$.

\subsection{Energy-momentum three-point function}

Regarding the three-point function of the energy-momentum tensor in flat space, it is determined by a triad of theory-dependent constants $\mathcal{A}$, $\mathcal{B}$, $\mathcal{C}$, the coefficients of three independent tensor structures $\mathcal{I}^{1,2,3}$~\cite{Osborn:1993cr,Erdmenger:1996yc}
\begin{multline}
S_d^3\,\langle T_{ab}(x_1) T_{cd}(x_2) T_{ef}(x_3)\rangle\\=
   \frac{\mathcal{A}\,\mathcal{I}^{(1)}_{abcdef}+\mathcal{B}\,\mathcal{I}^{(2)}_{abcdef}+\mathcal{C}\,\mathcal{I}^{(3)}_{abcdef}}{|x_{12}|^{d}\,|x_{23}|^{d}\,|x_{31}|^{d}}.
\end{multline}
Ward identities relating three- and two-point functions impose the following universal relation with $C_T$
\begin{equation}
    2d(d+2)C_T=(d-1)(d+2)\mathcal{A}-2\mathcal{B}-4(d+1)\mathcal{C}.
\end{equation}
In~\cite{Hofman:2008ar}, Hofman and Maldacena considered a thought experiment of a conformal collider. The expectation value of the energy flux measured far away in a state prepared by a localized insertion of the energy-momentum tensor in vacuum depends on two parameters, $t_2$ and $t_4$. The theory-dependent energy flux parameters are then related to the three-point function triad~\cite{Buchel:2009sk} in generic dimension $d$ as
\begin{multline}
    \frac{d^2(d+2)}{d+1}C_T\,t_2 = (d-2)(d+1)(d+2)\mathcal{A}+3d^2\mathcal{B}\\
    -4d(2d+1)\mathcal{C}
\end{multline}
and
\begin{multline}
    \frac{-2d^2(d+2)}{d+1}C_T\,t_4 = (d+2)(2d^2-3d-3)\mathcal{A}+2d^2(d+2)\mathcal{B}\\-4d(d+1)(d+2)\mathcal{C}.
\end{multline}
Holographically, the determination of the expectation value of the energy flux amounts to evaluating the bulk action on a perturbation of the AdS background: a shock wave dual to the flux operator itself and a metric fluctuation dual to the localized insertion of the energy-momentum operator on the vacuum state. It is then enough to keep track of terms linear in the shockwave deformation and quadratic in the metric fluctuation.  
Initially, in 4D, Hofman and Maldacena related the flux parameters $t_2$ and $t_4$ to the Weyl anomaly central charges $a$ and $c$. After extensions to higher-curvature gravities and to 6D and 8D by Chen and L\"u~\cite{Chen:2023bgj}, it became apparent that for even dimensions higher than four, the $a$-charge \textit{cannot} be written as a linear combination of the three-point function parameters alone. Moving to 6D CFTs, the Weyl anomaly features three type-B central charges ($c_1, c_2, c_3$) and Chen and L\"u explicitly calculated the holographic 6D Weyl anomaly and demonstrated that while the $a$-charge remains independent, \textit{all three $c$-charges} are entirely fixed as linear combinations of $\mathcal{A}$, $\mathcal{B}$, and $\mathcal{C}$.
Understanding these relations is important for characterizing the space of consistent CFTs in higher dimensions and to connect flat‑space observables to geometric responses. The present work develops this connection in full generality for even-dimensional CFTs.

We start in subsection~\ref{sec:HoloDerivation} by reviewing the holographic computation of $C_T$ and relating it to the bulk Newton constant, or, equivalently, the Planck length. In subsection~\ref{sec:HoloWeyl}, we discuss the holographic Weyl anomaly for Einstein gravity with a negative cosmological constant and its relation with the $Q$-curvature. In subsection~\ref{sec:WeylSquared}, we present the Chern--Gauss--Bonnet theorem on compact even-dimensional Einstein manifolds, as recently obtained by Case et al.~\cite{Case:2024oih}, which relates the Euler density to the $ Q$-curvature and pointwise Weyl invariants. In particular, we keep track of the coefficient $c_{2,1}$ of the quadratic term in the Weyl tensor. In subsection~\ref{sec:HoloData}, we combine the previous ingredients to obtain the universal relation between $C_T$ and $c$. In section~\ref{sec:UniversalRel}, we present two alternative derivations in CFT. In section~\ref{sec:Examples}, we compare the universal relation with several known examples.
We proceed in section~\ref{sec:UniversalRel2} with the holographic computation of the energy-momentum three-point function and the Weyl anomaly stemming from three specific bulk higher-derivative gravities. We establish universal relations between the parameters and illustrate them with a few worked examples. Finally, we conclude in section~\ref{sec:Conclusion}.

\section{Universal relation between \texorpdfstring{$C_T$}{C\_T} and \texorpdfstring{$c_{2,1}$}{c_{2,1}} from holographic data}
We first determine the precise relation by examining holographic data obtained from two distinct computations. On the one hand, the holographic computation of $C_T$ in bulk Einstein gravity involves the Anti-de~Sitter (AdS) background radius and the Planck length. On the other hand, we report the coefficient of the quadratic term in the Weyl tensor of the corresponding holographic Weyl anomaly, which is essentially given by the $Q$-curvature. We then make use of the Chern-Gauss--Bonnet theorem by Case et al.~\cite{Case:2024oih} on compact Einstein manifolds to pinpoint the precise relation between the $Q$-curvature and the Weyl invariant relevant for our analysis. This coefficient is expressed in terms of the AdS radius and the Planck length. Eliminating the two-dimensional quantities, we end up with the algebraic relation between the two coefficients of the CFT data.

\subsection{\label{sec:HoloDerivation}Holographic derivation of \texorpdfstring{$C_T$}{C\_T} from bulk Einstein gravity}

One of the early successes of the AdS/CFT calculational prescription was the holographic derivation of $C_T$ for the CFT dual to bulk Einstein gravity by Liu and Tseytlin~\cite{Liu:1998bu}. The (Euclidean) gravitational action is given by
\begin{multline}
    S_{\text{EH}}=-\frac{1}{2l_{\text{P}}^{d-1}}\int_{M} d^{d+1}x\sqrt{\hat{g}}\left(\hat{R}-2\hat{\Lambda}\right)\\-\frac{1}{l_{\text{P}}^{d-1}}\int_{\partial M} d^dx\sqrt{g}\left(K-\hat{\lambda}\right),
\end{multline}
where $l_{\text{P}}$ is the Planck length and the bulk cosmological constant can be written in terms of the radius of the vacuum AdS solution as $2\hat{\Lambda}=-d(d-1)/L^2$. The boundary terms comprise the standard Gibbons-Hawking-York term~\cite{Gibbons:1998zr} and an \emph{effective} boundary cosmological constant $\hat{\lambda}$ that is tuned later on to $(d-1)/L$ to ensure the conformal invariance of the action computed on the solution of the Dirichlet problem for the metric fluctuation. The quadratic part of the gravitational action, as a function of the boundary metric fluctuation $h$, turned out to be given by a non-local kernel that matches the CFT expectation for the two-point function of the energy-momentum tensor in flat space,
\begin{align}
    S^{(2)}_{\text{EH}}\,=\,&\frac{\Gamma(d+2)L^{d-1}}{8\pi^{d/2}(d-1)l_{\text{P}}^{d-1}}\nonumber\\\times&\int_{\mathbb{R}^d}d^{d}x\int_{\mathbb{R}^d}d^{d}y\frac{h_{ab}(x)\mathcal{I}_{ab,cd}(x-y)h_{cd}(y)}{|x-y|^{2d}}.
\end{align}
Performing the second functional variation with respect to the source $h$, one obtains the holographic two-point correlation function for the energy-momentum tensor and, with it, the $C_T$ coefficient 
\begin{equation}
    C_T=4\pi^{d/2}\frac{\Gamma(d+2)}{(d-1)\Gamma^3(d/2)}\frac{L^{d-1}}{l_{\text{P}}^{d-1}}.
\end{equation}

\subsection{\label{sec:HoloWeyl}Holographic Weyl anomaly for bulk Einstein gravity and \texorpdfstring{$Q$}{Q}-curvature}

As an outgrowth of the IR/UV connection~\cite{Susskind:1998dq}, bulk infrared (IR) divergences can be mapped to the ultraviolet (UV) divergences of the boundary theory. Thus, the Weyl anomaly of the dual CFT is determined from the bulk action. The detailed computation of this \textit{holographic} Weyl anomaly was originally carried out by Henningson and Skenderis~\cite{Henningson:1998ey,Henningson:1998gx}. This sparked interest among conformal geometers and led Graham and Zworski~\cite{Graham2003} to discover that the integrated holographic Weyl anomaly coincides with the integrated $Q$-curvature---a central object in Conformal Geometry.

A particular bulk metric, a solution to the Einstein equation, enables straightforward computations and lets us identify the numerical coefficients in front of the $Q$-curvature.  
Consider the particular Poincar\'e--Einstein metric $\hat{g}$ with an Einstein metric $g_{\text{\tiny E}}$ at its conformal infinity (PE/E)~\cite{Besse:1987pua}
\begin{equation}
    \hat{g}=L^2\frac{dx^2+(1-\lambda x^2/2)^2g_{\text{\tiny E}}}{x^2}.
\end{equation}
Here $\lambda=R/[2d(d-1)]$ is a multiple of the necessarily constant Ricci scalar of the boundary Einstein manifold.
The volume anomaly is given by the logarithmically divergent term of the regularized volume expansion
\begin{align}
    \Vol|_{x>\epsilon}&=L^{d+1}\Vol[g_{\text{\tiny E}}]\int_{\epsilon} \frac{dx}{x^{d+1}}(1-\lambda x^2/2)^d\nonumber\\&\sim L^{d+1}\Vol[g_{\text{\tiny E}}] \int_{\epsilon}\frac{dx}{x}(-\lambda/2)^{d/2}\binom{d}{d/2}.
\end{align}
The volume anomaly times the Einstein--Hilbert Lagrangian density on the PE/E metric (which evaluates to $d/[L^2l_{\text{P}}^{d-1}]$) gives the integrated holographic Weyl anomaly, so that one can read  
\begin{equation}\label{eq:HoloWeyl}
    (4\pi)^{d/2}\langle T\rangle=-\frac{4(-\pi)^{d/2}}{\Gamma^2(d/2)}\frac{L^{d-1}}{l_{\text{P}}^{d-1}}Q_d+\text{ttd},
\end{equation}
where we have used the fact that the critical $Q$-curvature, in our conventions,\footnote{In particular, we have $Q_d=(d-1)!$ on the standard round unit sphere.} reduces to $Q_d=(2\lambda)^{d/2}(d-1)!$ on a compact Einstein manifold. We need to establish the connection between the critical $Q$-curvature and the pointwise Weyl-invariant quadratic in the Weyl tensor.

\subsection{\label{sec:WeylSquared}Weyl-squared content of the \texorpdfstring{$Q$}{Q}-curvature}

A key result in~\cite{Case:2024oih} is that any scalar contraction of the Weyl tensor on an Einstein manifold can be completed with total derivative terms, powers of the Laplacian, thereby making it a pointwise conformal invariant.

In particular, when applied to the Pfaffian that is polynomial in the Weyl tensor, one ends up with an explicit formula for the Pfaffian in terms of pointwise conformal invariants and a pure curvature scalar power that corresponds to the critical $Q$-curvature.

For our purposes, it suffices to focus on the connection among $E_d$, $Q_d$, and $W^{(d)}_{2,1}$. We consider the following excerpt from Theorem~1.2 in~\cite{Case:2024oih},
\begin{equation}
    \Pf=(2\lambda)^{d/2}(d-1)!!+\frac{(-2)^{2-d/2}}{(d/2-1)!}\frac{1}{8}W^{(d)}_{2,1}+\text{lot}+\text{ttd}.
\end{equation} 
The convention for the Pfaffian $\Pf$ is that the volume integral on a compact $d$-manifold produces $(2\pi)^{d/2}$ times the Euler characteristic. Therefore, to translate this to our convention for the Euler density, we first note that
\begin{equation}
    E_d=\frac{(-1)^{d/2}}{2^{d/2}}\epsilon_d\epsilon_d\Riem\cdots\Riem\qquad
\end{equation}
and
\begin{eqnarray}
 \Pf&=&\frac{(d-1)!!}{2^{d/2}d!}\epsilon_d\epsilon_d\Riem\cdots\Riem\nonumber\\
 &=&\dfrac{(-1)^{d/2}(d-1)!!}{d!}E_d.
\end{eqnarray}
Second, we recall that the critical $Q$-curvature on a compact Einstein space is given by a multiple of the appropriate power of the constant scalar curvature,
\begin{equation}
    Q_d=(2\lambda)^{d/2}(d-1)!.
\end{equation}
Thus, the $Q$-curvature is related with our preferred quadratic in Weyl tensor $W^{(d)}_{2,1}$ as
\begin{equation}
    Q_d=\frac{(-1)^{d/2}}{d}\left(E_d-\frac{d}{4}W^{(d)}_{2,1}+\text{lot}+\text{ttd}\right).
\end{equation} 
Note that this convention for pointwise Weyl invariance differs slightly from that of Case et al.~\cite{Case:2024oih}. We reserve $W^{(d)}_{2,1}$ to denote the $d$-dimensional invariant induced by the corresponding ambient curvature invariant.

We can now write the holographic value of $c_{2,1}$ from Eq. \eqref{eq:HoloWeyl} 
\begin{equation}
    c_{2,1}=a=\frac{d\pi^{d/2}}{[(d/2)!]^2}\dfrac{L^{d-1}}{l_{\text{P}}^{d-1}}.
\end{equation}
Alternatively, we can also connect the $Q$-curvature with a term quadratic in the Weyl tensor by careful examination of its leading and subleading terms in curvature (see discussion around eq.~\eqref{eq:Sublead})
\begin{equation}
    Q_d=-\frac{d-2}{8(d-3)}W(-\nabla^2)^{d/2-2}W+\text{lot}+\text{ttd}.
\end{equation}

\subsection{\label{sec:HoloData}\texorpdfstring{$C_T$}{C\_T} versus \texorpdfstring{$c_{2,1}$}{c\_{2,1}} from holographic data}

We can now compare the two holographic results for $C_T$ and $c$ and eliminate the dependence on the AdS radius and Planck length to obtain 
\begin{equation}
    C_T=\frac{d}{d-1}\frac{(d+1)!}{(d/2-1)!}c_{2,1}
.\end{equation}
When written in terms of the $C_{TT}$ coefficient of the $TT$-correlator in momentum space (cf.~\cite{Bzowski:2017poo})
\begin{equation}
    \langle\!\langle T_{ab}(\boldsymbol{p})T_{cd}(-\boldsymbol{p})\rangle\!\rangle=C_{TT}\Pi_{ab,cd}(\boldsymbol{p})p^{d},
\end{equation}
with $\Pi_{ab,cd}$ being the transverse-traceless projector and the Dirac delta enforcing momentum conservation factored out, the above relation simplifies to
\begin{equation}
    (4\pi)^{d/2}C_{TT}=-\frac{(-1)^{d/2}d}{2}c_{2,1}.
\end{equation}

Without a genuine CFT derivation, the relation above might appear to be nothing more than a numerical coincidence. We therefore turn to the task of establishing this central relation by examining more closely the connection between the two-point function of the energy-momentum tensor and the trace anomaly in a generic even-dimensional CFT.

\section{\label{sec:UniversalRel} Universal relation between \texorpdfstring{$C_T$}{C\_T} and \texorpdfstring{$c_{2,1}$}{c\_{2,1}} within the CFT}

The CFT derivations exploit the fact that the trace anomaly controls the scale variation of the partition function $Z_{\text{CFT}}$ or effective action $W_{\text{CFT}}$ (see, e.g.,~\cite{Osborn:1993cr}) 
\begin{equation}
    \mu\frac{d}{d\mu}W_{\text{CFT}}=\int_{M}d\text{vol}_g\langle T\rangle.
\end{equation}
Upon functional differentiation with respect to the background metric, the running of the energy-momentum tensor two-point function on flat space from the LHS of the above relation reads
\begin{align}
    \mu\dfrac{d}{d\mu}\langle T_{ab}(x)T_{cd}(0)\rangle&=\frac{C_T}{S_d^2}\mu\dfrac{d}{d\mu}\frac{\mathcal{I}_{ab,cd}(x)}{|x|^{2d}}\nonumber\\&=\frac{C_T}{S_d^2}\frac{\Delta^T_{ab,cd}}{4(d-2)^2d(d+1)}\mu\frac{d}{d\mu}\frac{1}{|x|^{2d-4}}.
\end{align}
Here $\Delta^T_{ab,cd}$ stands for a quartic differential operator that ensures conservation and tracelessness, namely $\Delta^T_{ab,cd}=(S_{ac}S_{bd}+S_{ad}S_{bc})/2-S_{ab}S_{cd}/(d-1)$, with $S_{ab}=\partial_a\partial_b-\delta_{ab}\partial^2$.

The scale dependence of $1/|x|^{2d-4}$ comes from the regularization prescription. With the following two identities within \textit{differential regularization}~\cite{Dunne:1992ws},
\begin{subequations}
\begin{equation}\label{eq:DifRegA}
    \frac{1}{|x|^p}=\frac{1}{(p-2)(p-d)}\partial^2\frac{1}{|x|^{p-2}},
\end{equation}
\begin{equation}\label{eq:DifRegB}
    \mu\frac{d}{d\mu}\left(\mathcal{R}\frac{1}{|x|^d}\right)=S_d\delta^{(d)}(x),
\end{equation}
\end{subequations}
it is straightforward to iterate using eq.~\eqref{eq:DifRegA} to lower the exponent until it reaches $d$, and then use eq.~\eqref{eq:DifRegB} to produce a Dirac delta. The outcome of this iteration is
\begin{multline}
    \mu\frac{d}{d\mu}\langle T_{ab}(x)T_{cd}(0)\rangle\\=\frac{(d-1)(d/2-1)!}{(4\pi)^{d/2}(d+1)!}C_T\Delta^T_{ab,cd}(\partial^2)^{d/2-2}\delta^{(d)}(x).
\end{multline}
On the other hand, we need to turn to the kernel of the second metric variation of the trace anomaly. It suffices to keep track of the quadratic term in the Weyl tensor. Furthermore, the Euler term does not contribute. One can directly expand the quadratic term in Weyl up to second order in the metric perturbation around flat space, or trade it for the Euler term and $Q$-curvature modulo total derivatives. The $Q$-curvature is pure Ricci and can be easily varied in 4d and 6d, where explicit expressions for it are at hand. Cubic and higher-order terms in the Weyl tensor can safely be discarded when expanding around flat space in the computation of the two-point function, since their contribution to the effective action is cubic or higher in the metric perturbation.

It may prove instructive to note that the connection between $C_T$ and $c_{2,1}$ can also be obtained within \emph{dimensional regularization} by careful examination of the pole structure. On the $C_T$ side, we have that the kernel $1/|x|^{2d-4}$ has a pole, in a distributional sense (see, e.g.,~\cite{Gelfand1964}) given by
\begin{equation}
    \frac{1}{|x|^{2d-4}}\sim\dfrac{1}{\epsilon}\frac{(d-2)S_d}{2^{d-3}(d-3)!}(\partial^2)^{d/2-2}\delta^{(d)}(x),
\end{equation}
that must be canceled by the one-loop counterterm, which in turn is given by the trace anomaly (see, e.g.,~\cite{Duff:2020dqb} and eqs.~(21--26) therein for 2d and 4d). This same pole-matching led to the discovery of the GJMS operators in the two-point function of scalar operators in~\cite{Diaz:2008hy}. See also~\cite{Schwimmer:2019efk} for a thorough analysis of scalar operators.

\subsection{Standard derivation in \texorpdfstring{$4d$}{4d}}

In 4d, we only expand Weyl conformal gravity $W^2$ on the RHS to quadratic order in the metric perturbation around flat space~\cite{Osborn:1993cr}. It is enough to consider the linear variation of the Weyl tensor, which is given by
\begin{equation}
    W_{abcd}=2\mathcal{E}^{W}_{abcd,efgh}\partial_f\partial_gh_{eh}+O(h^2),
\end{equation}
where $\mathcal{E}^{W}_{abcd,efgh}$ denotes the projector onto the space of tensors with the same algebraic properties as the Weyl tensor, namely $\mathcal{E}^{W}_{abcd,efgh}=\partial W_{abcd}/\partial W_{efgh}$. After some algebraic manipulations, the RHS can be cast in the form
\begin{align}
    \mu\frac{d}{d\mu}\langle T_{ab}(x)T_{cd}(0)\rangle&=\frac{C_T}{(4\pi)^{d/2}40}\Delta^T_{ab,cd}\delta^{(4)}(x)\nonumber\\&=\dfrac{4c}{(4\pi)^2}\Delta^T_{ab,cd}\delta^{(4)}(x),
\end{align}
leading to the well-known (modulo conventions) relation
\begin{equation}
    C_T=160\,c\qquad(d=4).
\end{equation}

\subsection{Alternative derivation in \texorpdfstring{$4d$}{4d}}

In four dimensions, we can take advantage of the fact that, modulo the topological Euler density, one can trade Weyl-squared conformal gravity for Lanczos conformal gravity, which is pure Ricci and essentially the critical 4d $Q$-curvature, modulo a trivial total derivative, by virtue of the Gauss--Bonnet formula. Sen and Sinha~\cite{Sen:2014nfa} followed this approach, expanding the Ricci tensor and scalar in Cartesian coordinates as
\begin{subequations}
\begin{eqnarray}
     R_{ab}&=&-\frac{1}{2}(\partial^c\partial_a h_{bc}+\partial^c\partial_bh_{ac}-\partial^2 h_{ab}-\partial_a\partial_bh)\nonumber\\
     &+& O(h^2)
\end{eqnarray}
\begin{equation}
    R=-\partial^a\partial^bh_{ab}+\partial^2 h+O(h^2),
\end{equation}
\end{subequations}
to derive the second-order term in the metric variation around flat space for
\begin{equation} 
    W^2\sim2(\Ric^2-R^2/3)\sim-4Q_4,
\end{equation} 
obtaining the very same result as before. 

\subsection{Standard derivation in \texorpdfstring{$6d$}{6d}}

Here we need to expand around flat space the six-dimensional Weyl conformal gravity, given by $I_3=W\nabla^2W+\text{lot}+\text{ttd}$, to quadratic order in the metric fluctuation $h_{ab}$, resulting in~\cite{Osborn:2015rna}
\begin{align}
    \mu\dfrac{d}{d\mu}\langle T_{ab}(x)T_{cd}(0)\rangle&=\frac{C_T}{(4\pi)^{3}504}\Delta^T_{ab,cd}\partial^2\delta^{(6)}(x)\nonumber\\&=\frac{6c_3}{(4\pi)^3}\Delta^T_{ab,cd}\partial^2\delta^{(6)}(x),
\end{align}
leading to the known (modulo conventions) relation
\begin{equation}
    C_T=3024\,c_3\qquad(d=6).
\end{equation}

\subsection{Alternative derivation in \texorpdfstring{$6d$}{6d}}

The six-dimensional analog of Lanczos conformal gravity is given by the L\"u--Pang--Pope conformal gravity~\cite{Lu:2013hx}, written in terms of the Ricci tensor and its derivatives. This 6d conformal gravity is exactly the holographic Weyl anomaly of~\cite{Henningson:1998gx} and, after Graham and Zworski~\cite{Graham2003}, the 6d critical $Q$-curvature. It is enough to restrict attention to the term quadratic in curvature, so that, up to the Euler density, cubic in curvature terms and trivial total derivatives, we have
\begin{eqnarray}
  I_3 &\sim& W\nabla^2W\nonumber\\
  &\sim& 3\Ric\nabla^2\Ric-\frac{9}{10}R\nabla^2R  \\
  &\sim& 6Q_6\sim\frac{3}{2}W^{(6)}_{2,1}.\nonumber
\end{eqnarray}
After some algebraic manipulations and expansion of the Ricci tensor and scalar, one obtains, as expected, the very same relation as before. Sen and Sinha originally announced this 6d relation~\cite{Sen:2014nfa}. 

\subsection{Standard derivation in general even dimensions}

Let us now present both derivations in a generic even dimension. We need to expand the term quadratic in the Weyl tensor in the trace anomaly. Within our conventions, up to trivial total derivatives and higher-order terms in curvature, this corresponds to
\begin{equation}
    (4\pi)^{d/2}\langle T\rangle=c_{2,1}\frac{d(d-2)}{8(d-3)}W(\nabla^2)^{d/2-2}W+\text{lot}+\text{ttd}. 
\end{equation}
Conformal symmetry fixes the structure of the RHS except for a numerical coefficient $\#$ in front of the $c$ charge
\begin{widetext}
\begin{align}
  \mu\frac{d}{d\mu}\langle T_{ab}(x)T_{cd}(0)\rangle=\frac{(d-1)(d/2-1)!}{(4\pi)^{d/2}(d+1)!}C_T\Delta^T_{ab,cd}(\partial^2)^{d/2-2}\delta^{(d)}(x)=\#\frac{c_{2,1}}{(4\pi)^{d/2}}\Delta^T_{ab,cd}(\partial^2)^{d/2-2}\delta^{(d)}(x).\label{eq:TTVEP}
\end{align}
\end{widetext}

Let us take a shortcut to avoid gymnastics with the many indices. If we restrict ourselves to acting on the transverse-traceless part of the metric fluctuation $h$ and go to momentum space, we can employ a useful identity worked out by Erdmenger and Osborn~\cite{Erdmenger:1997wy}, together with the fact that $\mathcal{E}^{W}_{abcd,efgh}$ is a projector. Returning to position space, the quadratic expansion reads  
\begin{equation}
    W(\nabla^2)^{d/2-2}W=\frac{d-3}{d-2}h^{\text{tt}}_{ad}(\partial^2)^{d/2}h^{\text{tt}}_{ad}+O(h^3).
\end{equation}
Functional derivatives produce a factor of 2 and two more factors of 2 from the definition of the energy-momentum tensor, and the quartic operator $\Delta^T_{ab,cd}$ on transverse-traceless tensors produces $(\partial^2)^2$ times the identity. So, the missing numerical factor is $\#=d$. Comparing both sides of the equality, we arrive via the CFT at the universal relation
\begin{equation}
    C_T=\frac{d}{d-1}\frac{(d+1)!}{(d/2-1)!}c_{2,1}
.\end{equation}

\subsection{Alternative derivation in general even dimensions}

As for the alternative route, we can invoke the connection with the $Q$-curvature, a pure Ricci curvature invariant. The metric variation of the critical $Q$-curvature is the obstruction tensor, and its leading asymptotics is generally known (see, e.g.,~\cite{Graham2005})
\begin{widetext}
\begin{align}
    Q_d&=(-\nabla^2)^{d/2-1}\frac{R}{2(d-1)}-\frac{d-2}{8(d-3)}W(-\nabla^2)^{d/2-2}W+\text{lot}\nonumber\\&=(-\nabla^2)^{d/2-1}\frac{R}{2(d-1)}-\left[\frac{1}{2}\Ric(-\nabla^2)^{d/2-2}\Ric-\frac{d}{8(d-1)}(-\nabla^2)^{d/2-2}R\right]+\text{lot}.\label{eq:Sublead}
\end{align}
\end{widetext}
We can therefore trade the term quadratic in Weyl for the $Q$-curvature,
\begin{equation}
    (4\pi)^{d/2}\langle T\rangle=-c_{2,1}(-1)^{d/2}dQ_d+\text{lot}+\text{ttd}. 
\end{equation}
As mentioned above, the metric variation of the $Q$-curvature is the Fefferman--Graham obstruction tensor (the higher-dimensional analog of the 4d Bach tensor), so we only need the linearization of the obstruction tensor around flat space.

However, this line of reasoning yields a remarkable observation. The second metric variation of the $Q$-curvature, which can be thought of as a particular conformal gravity Lagrangian admitting Einstein solutions, is just the kinetic term of the corresponding \emph{Weyl graviton} when restricted to the transverse-traceless component of the metric fluctuation. This is nothing but the GJMS-like operator on symmetric transverse-traceless two-tensors that happens to factorize into shifted Lichnerowicz Laplacians on Einstein manifolds, as shown some time ago by Matsumoto~\cite{Matsumoto2013}. In particular, in flat space, it reduces to a power of the Laplacian. It is just a matter of keeping track of the overall coefficient to be able to obtain the numerical factor that fixes the RHS of eq.~\eqref{eq:TTVEP},
\begin{equation}
    Q_d=-\frac{1}{8}h^{\text{tt}}_{ad}(-\partial^2)^{d/2}h^{\text{tt}}_{ad}+O(h^3).
\end{equation}
This completes the alternative derivation \emph{\`a la Lanczos}, i.e., trading Weyl (or Riemann) tensors by pure Ricci terms in the conformal gravity Lagrangian. 

\section{\label{sec:Examples}Examples}

A vast literature exists on four- and six-dimensional CFTs (see, e.g.,~\cite{Beccaria:2015vaa,Beccaria:2015uta} and references therein). We will focus on the less-explored instances $d\geqslant8$.

For conformal powers of the Laplacian, GJMS operators, there was a conjectured value of $C_T$ up to the critical one (i.e., up to $(\partial^2)^{d/2}$ in flat space) due to Osborn and Stergiou~\cite{Osborn:2016bev}. In their original paper, the authors verified the results in dimensions 2, 4, and 6. The eight-dimensional case was confirmed shortly thereafter in~\cite{Guerrieri:2016whh}, providing some insight into a general proof, based on conformal partial wave (CPW) expansions of four-point functions for free fields, which they summed over conformal primaries. See table~\ref{tab:1}.

\begin{table*}[t]
\caption{\label{tab:1}The explicit connection between derivative order and $C_T$}
\begin{ruledtabular}
\begin{tabular}{cr}
Field & $C_T$ \\\colrule
$\partial^2\textrm{ conformal scalar}$ & $d/(d-1)$ \\
$\partial^4\textrm{ conformal scalar}$ & $-2d(d+4)/(d-1)(d-2)$ \\
$\partial^6\textrm{ conformal scalar}$ & $3d(d+4)(d+6)/(d-1)(d-2)(d-4)$ \\
$\partial^8\textrm{ conformal scalar}$ & $-4d(d+4)(d+6)(d+8)/(d-1)(d-2)(d-4)(d-6)$
\end{tabular}
\end{ruledtabular}
\end{table*}

\subsection{8d GJMS}

Taking advantage of the factorization properties of the GJMS operators on Einstein manifolds, we have been able to compute the accumulated relevant heat coefficient by using the explicit result of Avramidi~\cite{Avramidi:1990je} for the $b_8$ diagonal heat coefficient in 8d. We also holographically reproduced the boundary anomaly from the 9d heat coefficients of the dual bulk massive scalar~\cite{Aros2026}, extending the program initiated in~\cite{Bugini:2018def} for $5d/4d$ and $7d/6d$. The central charge $c_{2,1}$ for the 8d $P_{2k}$ operators turned out to be given by the following polynomial in $k$
\begin{equation}
    9!\,c_{2,1}=-\frac{k^9}{24}+\frac{3k^7}{4}-\frac{91k^5}{40}-\frac{41}{6}k^3+\frac{72k}{5}.
\end{equation}
Comparison with the expectation of the Osborn--Stergiou predictions can only be made for subcritical and critical $P_{2k}$, \textit{i.e.}, for $k=1,2,3,4$ in 8d. In Table~\ref {tab:2}, we show results that are in complete agreement with the previously derived universal relation between $C_T$ and $c_{2,1}$. 
\begin{table}[b]
\caption{\label{tab:2}Critical $C_T$ and $c_{2,1}$}
\begin{ruledtabular}
\begin{tabular}{cccc}
$k$ & Field & $C_T$ & $9!\,c_{2,1}$ \\\colrule
$1$ & $\partial^2\textrm{ conformal scalar}$ & $8/7$ & $6$ \\
$2$ & $\partial^4\textrm{ conformal scalar}$ & $-32/7$ & $-24$ \\
$3$ & $\partial^6\textrm{ conformal scalar}$ & $24$ & $126$ \\
$4$ & $\partial^8\textrm{ conformal scalar}$ & $-256$ & $-1344$ \\
\end{tabular}
\end{ruledtabular}
\end{table}
 
\subsection{Holographic higher-order gravities}

We can also add to the list of examples, of course, the holographic data for bulk Einstein gravity, which we initially used to obtain the numerical relation
\begin{align}
C^\text{E}_T&=4\pi^{d/2}\frac{\Gamma(d+2)}{(d-1)\Gamma^3(d/2)}\frac{L^{d-1}}{l_{\text{P}}^{d-1}},\\c_{2,1}^\text{E}&=\frac{4\pi^{d/2}}{d\Gamma^2(d/2)}\frac{L^{d-1}}{l_{\text{P}}^{d-1}}.
\end{align}
Actually, we can do better and consider adding to the Einstein--Hilbert Lagrangian any (local) higher curvature term, i.e., contractions of the Riemann tensor, including its covariant derivatives as well, as
\begin{multline}
    S_{\text{G}}[\hat{g}]=-\frac{1}{2l_{\text{P}}^{d-1}}\int_{M} d^{d+1}x\sqrt{\hat{g}}\Bigg(\frac{d(d-1)}{L^2}+\hat{R}\\+\sum_{n\geqslant2}\alpha_n L^{2n-2}\hat{I}_{n}\Bigg).
\end{multline}
Here, $L$ is a length scale introduced to make the couplings $\alpha_n$ dimensionless. The (Euclidean) AdS vacuum of the original Einstein--Hilbert action has radius $L$, but the putative AdS vacuum of the higher-derivative theory will generally have an effective radius. At linear order in the couplings $\alpha_n$, such a solution is guaranteed to exist. In general, one must find a positive, typically nondegenerate root of an algebraic equation for the ratio between the length scale $L$ and the effective AdS radius $\tilde{L}=L/\sqrt{f_\infty}$.

Let us follow a shortcut route~\cite{Hawking:1976ja,Duff:1993wm,Bugini:2016nvn} to get this algebraic equation. This is equivalent to extremizing the gravitational action with respect to the AdS radius as in~\cite{Beccaria:2015ypa}. We exploit the scale invariance of the action to write 
\begin{multline}
    S_{\text{G}}[\mu^2\hat{g}]=-\frac{\mu^{d+1}}{2l_{\text{P}}^{d-1}}\int_{M} d^{d+1}x\sqrt{\hat{g}}\Bigg(\dfrac{d(d-1)}{L^2}+\dfrac{\hat{R}}{\mu^2}\\+\sum_{n\geqslant2}\alpha_n L^{2n-2}\frac{\hat{I}_{n}}{\mu^{2n}}\Bigg), 
\end{multline}
and then take
\begin{equation}
    \frac{d}{d\mu}S_{\text{G}}[\mu^2\hat{g}]\bigg|_{\mu=1}=0.
\end{equation}
It immediately follows from dimensional arguments that, on the AdS vacuum, the Lagrangian must depend on the ratio $f_\infty/\mu^2$, as
\begin{equation}
    S_{\text{G}}[\mu^2\hat{g}_{\text{\tiny AdS}}]=\mu^{d+1}\int_{M} d^{d+1}x\sqrt{\hat{g}_{\text{\tiny AdS}}}\mathcal{L}\left(\frac{f_\infty}{\mu^2}\right),
\end{equation}
so that the scale invariance condition translates into 
\begin{equation}
   (d+1)\mathcal{L}(f_\infty)-2f_\infty{\cal L}'(f_{\infty})=0.
\end{equation}
It is convenient to rewrite this algebraic equation in terms of the auxiliary function 
\begin{equation}
    h(f_{\infty})=\frac{-2l_{\text{P}}^{d-1}L^2}{d(d-1)}\left(\mathcal{L}(f_\infty)-\frac{2f_\infty}{d+1}\mathcal{L}'(f_\infty)\right),
\end{equation}
that satisfies $h(f_{\infty})=0$ on-shell and, after a suitable rescaling of the invariants, has the simple expansion 
\begin{equation}
    h(f_{\infty})=1-f_\infty+\sum_{n\geqslant2}\alpha_n f_\infty^n. 
\end{equation}
We can now apply our \emph{simple holographic recipe}~\cite{Bugini:2016nvn,Aros2026} to determine the holographic central charges $a$ and $c_{2,1}$ by evaluating the action on the particular Poincar\'e--Einstein bulk metric with an Einstein metric at its conformal infinity. From the pure-Ricci part, we get the action at AdS precisely, while from the deviations, which involve the Weyl tensor, it is enough to keep the Weyl-squared term 
\begin{multline}
    S_{\text{G}}[\hat{g}_{\text{\tiny PE/E}}]=\mathcal{L}(f_\infty)\int_{M}d^{d+1}x\sqrt{\hat{g}_{\text{\tiny PE/E}}}\hat{1}\\+\int_{M}d^{d+1}x\sqrt{\hat{g}_{\text{\tiny PE/E}}}\gamma\hat{W}^2+\cdots.
\end{multline}
From the first term, $\hat{1}$, we obtain the volume anomaly, proportional to the critical $Q$-curvature $Q_d$. By completing the quadratic $\hat{W}^2$ term with appropriate derivatives, we obtain the specific conformal invariant $\hat{W}_{2,1}^{(d+1)}$ on Einstein manifolds, which, as derived by Case et al.~\cite{Case:2024oih}, is quadratic in the bulk Weyl tensor and induces the corresponding Weyl invariant $W^{(d)}_{2,1}$ on the Einstein boundary. Keeping track of the numerical coefficients, we obtain for the holographic Weyl anomaly
\begin{widetext}
\begin{align}
    (4\pi)^{d/2}\langle T\rangle&=-(-1)^{d/2}adQ_d+(c_{2,1}-a)\frac{d}{4}W^{(d)}_{2,1}+\text{lot}+\text{ttd}\nonumber\\&=-(-1)^{d/2}\frac{\pi^{d/2}\tilde{L}^{d+1}}{[(d/2)!]^2}\mathcal{L}(f_\infty)dQ_d-\frac{(2\pi)^{d/2}\tilde{L}^{d-3}}{(d/2)!(d-4)!!}8\gamma\frac{d}{4}W^{(d)}_{2,1}+\text{lot}+\text{ttd}.    
\end{align}
\end{widetext}
In all, the holographic central charges $a$ and $c_{2,1}$ are given by
\begin{align}
    a&=\frac{\pi^{d/2}\tilde{L}^{d+1}}{[(d/2)!]^2}\mathcal{L}(f_\infty)\label{eq:HoloCentral}\\
    c_{2,1}-a&=-\frac{8d(d-2)\pi^{d/2}\tilde{L}^{d-3}}{[(d/2)!]^2}\gamma
\end{align}
This result for the central charge $a$ is well known, and was first obtained by Imbimbo et al.~\cite{Imbimbo:1999bj}. As for the $c_{2,1}$ charge, in the particular case of \emph{Einstein‑like gravities}---though this still comprises a large family of gravitational theories (cf.~\cite{Bueno:2018yzo})---the structure of the coefficient $\gamma$ admits further refinement,
\begin{equation}
    L^4\mathcal{L}''(f_\infty)=2(d+1)d(d-1)(d-2)\gamma,
\end{equation} 
and, after basic algebraic steps, we derive an expression for $c$ that tweaks the canonical Einstein--Hilbert value in a notably simple way,
\begin{equation}
    c_{2,1}=-h'(f_\infty)\frac{d\pi^{d/2}\tilde{L}^{d-1}}{[(d/2)!]^2l_{\text{P}}^{d-1}}=-h'(f_\infty)c^{\text{E}}\Big|_{\tilde{L}}.
\end{equation}
Remarkably, using our universal relation, we can confirm the corresponding value for $C_T$ reported in~\cite{Bueno:2018yzo},
\begin{eqnarray}
    C_T &=&-h'(f_\infty)\frac{4\pi^{d/2}(d+1)!\tilde{L}^{d-1}}{(d-1)[(d/2-1)!]^3l_{\text{P}}^{d-1}} \nonumber\\
    &=&-h'(f_\infty) C_T^{\text{E}}\Big|_{\tilde{L}}.
\end{eqnarray}

Let us finish this discussion by recalling the observation of Li, L\"u and Mai~\cite{Li:2018drw} on $a$ and $c_{2,1}$. Provided $a$ and $c_{2,1}$ are written in terms of the effective AdS radius $\tilde{L}$ \emph{on-shell}, i.e., with $L=L(\tilde{L})$ and $f_\infty$ being a mere number determined by the couplings of the theory, then it is satisfied that
\begin{equation}
    c_{2,1}=\frac{1}{d-1}\tilde{L}\dfrac{\partial}{\partial\tilde{L}}a.
\end{equation}
This can be readily verified by taking the (log-)derivative of the holographic central charge $a$ in eq.~\eqref{eq:HoloCentral},
\begin{align}
    \tilde{L}\frac{\partial}{\partial\tilde{L}}a&=\frac{2\pi^{d/2}L^{d+1}}{(d+1)[(d/2)!]^2}\left(-2f_\infty\frac{\partial}{\partial f_\infty}\right)\frac{\mathcal{L}'(f_\infty)}{f_\infty^{(d-1)/2}}\nonumber\\&=-h'(f_\infty)\frac{d(d-1)\pi^{d/2}\tilde{L}^{d-1}}{[(d/2)!]^2l_{\text{P}}^{d-1}}\nonumber\\&=-h'(f_\infty)(d-1)c_{2,1}^{\text{E}}\Big|_{\tilde{L}}=(d-1)c_{2,1},
\end{align}
where we have used the on-shell condition to trade $\mathcal{L}(f_\infty)$ by $2 f_\infty\mathcal{L}'(f_\infty)/(d+1)$ and the fact that the latter only depends on $\tilde{L}$, as opposed to $\mathcal{L}(f_\infty)$ which also depends on $L$. This enables one to take derivatives with respect to $f_{\infty}$. Using the present universal relation, we end up with a prescription to compute $C_T$ directly from the type-A central charge $a$ in Einstein-like gravities. A comparison with~\cite{Bueno:2018yzo} (eq.~(18) and footnote~[90] therein) is also in order.

\section{\label{sec:UniversalRel2} Universal relation between \texorpdfstring{$\langle TTT\rangle$}{<TTT>} and Weyl anomaly from holographic data}

Now we turn to the energy-momentum three-point function and holographically compute the flux parameters and the Weyl anomaly to establish the relation.
It will prove enough to turn on three bulk curvature invariants, one quadratic in the Weyl tensor and two cubic. They will not modify the putative AdS background radius. Let us consider the Einstein--Hilbert action with the addition of the following four- and six-derivative curvature invariants

\begin{multline}
    S_{bulk}=-\frac{1}{2l_{\text{P}}^{d-1}}\int_{M} d^{d+1}x\sqrt{\hat{g}}\left\{\hat{R}-2\hat{\Lambda} \right.\\
    \left.+\gamma\, \hat{W}^2 + \kappa\, \hat{W}^3 + \delta\, \hat{W}'^3\right\}.
\end{multline}
\subsection{\label{sec:HoloDerivationThree}Three-point function from bulk gravity}
We now determine $C_T$ and the energy flux parameters $t_2$ and $t_4$ from the above bulk Lagrangian following the general result by Chen and L\"u~\cite{Chen:2023bgj}, (eqns. 51-52 therein). We first have to express the Weyl-squared part in terms of the Ricci scalar and Ricci and Riemann tensors squared, to read off their coefficients $e_{2,1}, e_{2,2}$ and $e_{2,3}$ 
\begin{equation*}
    \hat{W}^2=\hat{R}iem^2-\dfrac{4}{d-1}\hat{R}ic^2+\dfrac{2}{d(d-1)}\hat{R}^2
\end{equation*}
The same has to be done with the cubic part, to read off the corresponding coefficients $e_{3,1}, e_{3,2}, e_{3,3}, e_{3,4}, e_{3,5}, e_{3,6}, e_{3,7} $ and $e_{3,8}$ 
\begin{align*}
    \hat{W}^3=\hat{R}iem^3&-\dfrac{12}{d-1}\hat{R}ic\hat{R}iem^2+\dfrac{6}{d(d-1)}\hat{R}\hat{R}iem^2\\
    &+\dfrac{24}{(d-1)^2}\hat{R}iem\hat{R}ic^2+\dfrac{16d}{(d-1)^3}\hat{R}ic^3\\ 
    &-\dfrac{24(2d-1)}{d(d-1)^3}\hat{R}\hat{R}ic^2+\dfrac{8(2d-1)}{d^2(d-1)^3}\hat{R}^3,
\end{align*}
\begin{align*}
    \hat{W}'^3=\hat{R}iem'^3&-\dfrac{3}{d-1}\hat{R}ic\hat{R}iem^2+\dfrac{3}{2d(d-1)}\hat{R}\hat{R}iem^2\\
    &+\dfrac{3(d+1)}{(d-1)^2}\hat{R}iem\hat{R}ic^2+\dfrac{2(3d-1)}{(d-1)^3}\hat{R}ic^3\\ 
    &-\dfrac{3(d^2+3d-2)}{d(d-1)^3}\hat{R}\hat{R}ic^2+\dfrac{d^2+3d-2}{d^2(d-1)^3}\hat{R}^3.
\end{align*}
In our conventions, the holographic results become
\begin{align*}
    C_T&=4\pi^{d/2}\frac{\Gamma(d+2)}{(d-1)\Gamma^3(d/2)}\frac{L^{d-1}}{l_{\text{P}}^{d-1}}\left\{1+\frac{4(d-2)\gamma}{L^2}\right\},\\\nonumber\\ 
    C_T\,t_2&=d(d-1)L^{d-1}\left\{\frac{4\gamma}{L^2}-\frac{12(3d+4)\kappa}{L^4}-\frac{3(7d+4)\delta}{L^4}\right\},\\\nonumber\\
    C_T\,t_4&=6d(d-1)(d+1)(d+2)L^{d-1}\left\{\frac{2\kappa}{L^4}+\frac{\delta}{L^4}\right\}.
\end{align*}

\subsection{\label{sec:HoloWeyl2}Holographic Weyl anomaly}

From the same bulk Lagrangian, we can also read off the holographic Weyl anomaly by following the simple recipe envisaged in~\cite{Bugini:2016nvn,Aros2026}. We evaluate in a bulk Poincaré-Einstein metric with an Einstein metric on its conformal boundary; the pure Ricci terms factorize and contribute as a factor to the volume anomaly given by the boundary critical Q-curvature. The bulk Weyl-squared term must be completed with powers of the Laplacian and the AdS radius to produce the local Weyl covariant combination $\hat{W}_{2,1}$. In contrast, the Weyl-cubed term produces the local Weyl covariants $\hat{W}_{3,1}$ and $\hat{W}_{3,2}$ instead. They descend to the corresponding pointwise Weyl invariants $W_{2,1}$, $W_{3,1}$, and $W_{3,2}$, respectively, in the shifted basis for the Weyl anomaly

\begin{align*}
    (4\pi)^{d/2}\langle T\rangle&=-(-1)^{d/2}adQ_d+(c_{2,1}-a)\frac{d}{4}W^{(d)}_{2,1}\\
    &-(c_{3,1}-a)\frac{2d}{3}W^{(d)}_{3,1}+(c_{3,2}-a)\frac{4d}{3}W^{(d)}_{3,2}\\
    &+\text{lot}+\text{ttd}. 
\end{align*}
It is then just a matter of bookkeeping to work out the holographic central charges,
\begin{align*}
    a&=\dfrac{d\pi^{d/2}L^{d-1}}{[(d/2)!]^2\,l_P^{d-1}},\\\nonumber
    \\
    c_{2,1}&=a\left\{1+\frac{4(d-2)\gamma}{L^2}\right\},\\\nonumber\\ 
    c_{3,1}&=a\left\{1-\frac{6(d-2)(d-4)\kappa}{L^4}\right\},\\\nonumber\\
    c_{3,2}&=a\left\{1+\frac{3(d-2)(d-4)\delta}{L^4}\right\}.
\end{align*}
The holographic results confirm that the AdS radius remains unchanged, and that for even $D>4$ the central charge $a$ receives no correction from the Weyl-squared or the Weyl-cube bulk terms.

\subsection{\label{sec:HoloData2} Universal relations from holographic data}

We can now compare the two holographic results and eliminate the dependence on the bulk couplings. First, we confirm the result obtained from bulk Einstein gravity for $C_T$, but this time including the Weyl-squared addition
\begin{equation}
    C_T=\frac{d}{d-1}\frac{(d+1)!}{(d/2-1)!}c_{2,1}.
\end{equation}
Second, we obtain for the flux parameters
\begin{align}
    t_2=&\dfrac{d(d-1)}{(d-2)(d-4)}\times \nonumber\\
    &\nonumber\\
    &\dfrac{(d-4)c_{2,1}+2(3d+4)c_{3,1}-(7d+4)c_{3,2}}{c_{2,1}},
\end{align}
and
\begin{equation}
    t_4=\dfrac{2d(d-1)(d+1)(d+2)}{(d-2)(d-4)}\,\dfrac{c_{3,2}-c_{3,1}}{c_{2,1}}.
\end{equation}
Alternatively, we can solve for the central charges in terms of the triad of CFT energy-momentum three-point function parameters
\begin{align}
\dfrac{2d^2}{d-1}\,\dfrac{(d+2)!}{(d/2-1)!}\,c_{2,1}=(d-1)(d+2){\cal A} -2{\cal B} -4(d+1){\cal C},\\
\nonumber
\end{align}
\begin{align}
4d^4\,\dfrac{(d+2)!}{(d/2-1)!}\,c_{3,1}&=-(2d^5-14d^4+19d^3-5d^2-6d+8){\cal A}\nonumber\\
\nonumber\\
&+2d^2(d^2-6d+6){\cal B} -4d^3(d+1){\cal C},
\end{align}
\begin{align}
2d^4\,\dfrac{(d+2)!}{(d/2-1)!}\,c_{3,2}&=-(d^5-6d^4+2d^3+13d^2-6d-8){\cal A}\nonumber\\
\nonumber\\
&-2d^2{\cal B} -4d(d+1)(3d-4){\cal C}.
\end{align}

\subsection{\label{sec:ExamplesII}Examples}

\subsubsection*{6D}
Let us consider the first nontrivial dimension, $d=6$. We obtain,
\begin{equation}
    C_T=3024\,c_{2,1},
\end{equation}
which agrees with~\cite{Beccaria:2015ypa}. For the flux parameters, we obtain 
\begin{align}
    t_2=\dfrac{15}{2}\,\dfrac{c_{2,1}+22c_{3,1}-23c_{3,2}}{c_{2,1}},\\
    \nonumber
\end{align}
and
\begin{equation}
    t_4=420\,\dfrac{c_{3,2}-c_{3,1}}{c_{2,1}}.
\end{equation}
We need a dictionary to relate our Weyl invariants to those of the standard 6D basis $I_1,I_2,I_3$.
Modulo trivial total derivatives, we have
\begin{align*}
    \dfrac{1}{2}W^{(6)}_{2,1}&=\Phi_6+4I_1-I_2,\\
    3\Phi_6&=I_3-16I_1+4I_2
\end{align*}
in terms of the Fefferman-Graham invariant $\Phi_6$, so that we get  
\begin{align*}
    I_1&=-W^{(6)}_{3,2}\;,\; I_2=W^{(6)}_{3,1}\;,\,I_3=\dfrac{3}{2}W^{(6)}_{2,1}-4W^{(6)}_{3,2}-W^{(6)}_{3,1}.
\end{align*}
and for the respective coefficients we get
\begin{align}
    c_{2,1}=c_3\;,\; 4c_{3,1}=c_3-c_2\;,\,8c_{3,2}=-4c_3-c_1.
\end{align}
When we substitute back into our relation, we obtain in 6D
\begin{align}
    t_2=\dfrac{15}{16}\,\dfrac{23c_1-44c_2+144c_3}{c_3},\\
    \nonumber
\end{align}
and
\begin{equation}
    t_4=-\dfrac{105}{2}\,\dfrac{c_1-2c_2+6c_3}{c_3}.
\end{equation}
These very same relations were obtained in 6D by Osborn and Stergiou (\cite{Osborn:2015rna}, eqn.5.22 therein) from the available data for the free scalar, free Dirac spinor, and free two-form field in~\cite{deBoer:2009pn}. By combining the central charges with the three-point function parameters, they were able to compute the numerical coefficients above.

\subsubsection*{Free scalar}
For the free scalar, Osborn and Petkou~\cite{Osborn:1993cr} and Erdmenger and Osborn~\cite{Erdmenger:1997wy} were able to determine the three parameters of the energy-momentum three-point function
\begin{equation}
    \dfrac{4(d-1)^3}{d^2} \left[{\cal A}, {\cal B}, {\cal C} \right]=\left[4d, -4d(d-2), -(d-2)^2\right].
\end{equation}
The same information is encoded in 
\begin{equation}
    C_T=\dfrac{d}{d-1}\;,\; t_2=0\;,\; t_4=\dfrac{(d-1)(d+1)}{2}.
\end{equation}
By making use of the universal relations we have uncovered, we end up with a prediction for the central charges in generic even dimension $d\geq 6$
\begin{align*}
c_{2,1}&=\dfrac{(d/2-1)!}{(d+1)!},\\
c_{3,1}&=-\dfrac{3d^2-18d-8}{4d(d+2)}\cdot\dfrac{(d/2-1)!}{(d+1)!}\\
c_{3,2}&=-\dfrac{d^2-6d-8}{2d(d+2)}\cdot\dfrac{(d/2-1)!}{(d+1)!}.
\end{align*}
These values should match the central charges obtained via heat coefficients. In 6D, we need to use the dictionary again to relate our Weyl invariants with the traditional ones, but after straightforward algebra we can verify the full agreement for the central charges~\cite{Bastianelli:2000hi}
\begin{equation}
    7! \left[c_1\,,\, c_2\,,\, c_3\right]=\left[-28/3\,,\, 5/3\,,\, 2\right],
\end{equation}
\begin{equation}
    7! \left[c_{2,1}\,,\, c_{3,1}\,,\, c_{3,2}\right]=\left[2\,,\, 1/12\,,\, 1/6\right].
\end{equation}

In \cite{Aros2026}, using Avramidi's heat coefficient $b_8$ (a.k.a. $a_4$) and computer-aided methods, the central charges for the 8D free scalar were computed. It is worth mentioning that the same results can be obtained via the heat kernel independently following  
\begin{equation}
    9! \left[c_{2,1}\,,\, c_{3,1}\,,\, c_{3,2}\right]=\left[6\,,\, -3/4\,,\, -3/10\right].
\end{equation}

\section{\label{sec:Conclusion}Conclusion}

We have established the universal relationship between the CFT coefficients $\mathcal{A}$, $\mathcal{B}$, $\mathcal{C}$ or $C_T$, $t_2$, $t_4$ and the central charges $c_{2,1}$, $c_{3,1}$, $c_{3,2}$ in arbitrary even dimensions $d\geqslant6$. This highlights the nontrivial connection between conformal field theories in flat space and the trace anomaly arising from gravitational coupling. For $C_T$ and $c_{2,1}$, perhaps the most concise---and most striking---formulation of the universal relation is in terms of the coefficient $C_{TT}$ of the momentum-space $TT$-correlator (as flat as it gets) and the critical $Q$-curvature $Q_d$ (as curved as it is)
 \begin{equation}
    \langle T\rangle=2\,C_{TT}\,Q_d+\text{lot}+\text{ttd}.
\end{equation}
This effectively extends previously known results~\cite{Osborn:1993cr, Erdmenger:1997wy,Sen:2014nfa,Osborn:2016bev} to higher, even dimensions and, at the same time, contains the Weyl-completion of the term quadratic in the Weyl tensor (cf.~\cite{Bzowski:2017poo}, eq.~(3.17) therein)\footnote{We thank P. McFadden for pointing out the suitability of the $C_{TT}$ coefficient, as opposed to the more conventional $C_T$.}.

It is noteworthy that several other contexts have established a similar connection to $C_T$. For example, R\'enyi entropy across a spherical entangling surface~\cite{Perlmutter:2013gua}, free energy on conically deformed spheres~\cite{Beccaria:2017lcz}, free energy on squashed spheres~\cite{Bueno:2018yzo}, and pseudoentropy for (slightly) deformed spheres in dS/CFT~\cite{Anastasiou:2025rvz}.
A direct mapping between these quantities and the Weyl anomaly coefficients appears compelling.

We have thereby addressed the question raised in~\cite{Chen:2023bgj} concerning the proliferation of $c$-charges as $d$ increases and their relation to $\mathcal{A}$, $\mathcal{B}$, $\mathcal{C}$. Despite the ever-increasing number of pointwise Weyl invariants in the anomaly, the energy-momentum three-point function is solely determined by the three terms containing the lowest powers of the Weyl tensor. The explicit dimension dependence allows us to explore non-unitary CFTs and holographic setups, and it also predicts the heat-kernel coefficients for free fields. We verified a few explicit instances. 

We leave for future work a genuine CFT derivation of the $t_{2}$ and $t_{4}$ relations by analyzing the renormalization-group running of the energy-momentum three-point function with the mass scale. The main technical difficulty lies in expanding the pointwise Weyl invariants to cubic order in the metric perturbation, a challenge that we expect to tackle using computer-aided symbolic computation. 

\begin{acknowledgments}
We thank G. Anastasiou, J.S. Case, F. D\'iaz, M.J. Duff, P. McFadden, A.A. Tseytlin, and O. Zanusso for useful discussions, clarifications, and pointing out relevant references. D.E.D. also thanks the faculty and staff of the Science Department at Valencia College for their warm welcome and support. This work was partially funded through FONDECYT-Chile 1220335. R.A. wishes to thank professors J. Gomis (Universitat de Barcelona), E. Bergshoeff (University of Groningen), M. Romo and H. Adami (Shanghai Institute for Mathematics and Interdisciplinary Sciences), and E. Joung (Kyung Hee University) for their kind hospitality during the completion of this work.
\end{acknowledgments}

\emph{Note added.} After completion of the first version of this work, we were informed by P. McFadden that in~\cite{Bzowski:2017poo} 
the authors carried out the equivalent of our \emph{standard CFT derivation} in momentum space and were able to relate their $C_{TT}$ to the coefficient of the $W(\nabla^2)^{d/2-2}W$ term in the trace anomaly. Their result is correct in 4d, but the appropriate extension to higher even dimensions requires an additional dimensional factor $(d-2)/[2(d-3)]$ that was overlooked. Nevertheless, we find their relation very instructive and, consequently, we have cast our own result in terms of $C_{TT}$, absorbing the dimensional factor into the critical $Q$‑curvature $Q_d$, thus giving a new turn to the universal relation.


\begin{thebibliography}{54}%
\makeatletter
\providecommand \@ifxundefined [1]{%
 \@ifx{#1\undefined}
}%
\providecommand \@ifnum [1]{%
 \ifnum #1\expandafter \@firstoftwo
 \else \expandafter \@secondoftwo
 \fi
}%
\providecommand \@ifx [1]{%
 \ifx #1\expandafter \@firstoftwo
 \else \expandafter \@secondoftwo
 \fi
}%
\providecommand \natexlab [1]{#1}%
\providecommand \enquote  [1]{``#1''}%
\providecommand \bibnamefont  [1]{#1}%
\providecommand \bibfnamefont [1]{#1}%
\providecommand \citenamefont [1]{#1}%
\providecommand \href@noop [0]{\@secondoftwo}%
\providecommand \href [0]{\begingroup \@sanitize@url \@href}%
\providecommand \@href[1]{\@@startlink{#1}\@@href}%
\providecommand \@@href[1]{\endgroup#1\@@endlink}%
\providecommand \@sanitize@url [0]{\catcode `\\12\catcode `\$12\catcode
  `\&12\catcode `\#12\catcode `\^12\catcode `\_12\catcode `\%12\relax}%
\providecommand \@@startlink[1]{}%
\providecommand \@@endlink[0]{}%
\providecommand \url  [0]{\begingroup\@sanitize@url \@url }%
\providecommand \@url [1]{\endgroup\@href {#1}{\urlprefix }}%
\providecommand \urlprefix  [0]{URL }%
\providecommand \Eprint [0]{\href }%
\providecommand \doibase [0]{https://doi.org/}%
\providecommand \selectlanguage [0]{\@gobble}%
\providecommand \bibinfo  [0]{\@secondoftwo}%
\providecommand \bibfield  [0]{\@secondoftwo}%
\providecommand \translation [1]{[#1]}%
\providecommand \BibitemOpen [0]{}%
\providecommand \bibitemStop [0]{}%
\providecommand \bibitemNoStop [0]{.\EOS\space}%
\providecommand \EOS [0]{\spacefactor3000\relax}%
\providecommand \BibitemShut  [1]{\csname bibitem#1\endcsname}%
\let\auto@bib@innerbib\@empty
\bibitem [{\citenamefont {Capper}\ and\ \citenamefont
  {Duff}(1974)}]{Capper:1974ic}%
  \BibitemOpen
  \bibfield  {author} {\bibinfo {author} {\bibfnamefont {D.~M.}\ \bibnamefont
  {Capper}}\ and\ \bibinfo {author} {\bibfnamefont {M.~J.}\ \bibnamefont
  {Duff}},\ }\bibfield  {title} {\bibinfo {title} {{Trace anomalies in
  dimensional regularization}},\ }\href {https://doi.org/10.1007/BF02748300}
  {\bibfield  {journal} {\bibinfo  {journal} {Nuovo Cim. A}\ }\textbf {\bibinfo
  {volume} {23}},\ \bibinfo {pages} {173} (\bibinfo {year} {1974})}\BibitemShut
  {NoStop}%
\bibitem [{\citenamefont {Deser}\ and\ \citenamefont
  {Schwimmer}(1993)}]{Deser:1993yx}%
  \BibitemOpen
  \bibfield  {author} {\bibinfo {author} {\bibfnamefont {S.}~\bibnamefont
  {Deser}}\ and\ \bibinfo {author} {\bibfnamefont {A.}~\bibnamefont
  {Schwimmer}},\ }\bibfield  {title} {\bibinfo {title} {{Geometric
  classification of conformal anomalies in arbitrary dimensions}},\ }\href
  {https://doi.org/10.1016/0370-2693(93)90934-A} {\bibfield  {journal}
  {\bibinfo  {journal} {Phys. Lett. B}\ }\textbf {\bibinfo {volume} {309}},\
  \bibinfo {pages} {279} (\bibinfo {year} {1993})},\ \Eprint
  {https://arxiv.org/abs/hep-th/9302047} {arXiv:hep-th/9302047} \BibitemShut
  {NoStop}%
\bibitem [{\citenamefont {Boulanger}(2007{\natexlab{a}})}]{Boulanger:2007st}%
  \BibitemOpen
  \bibfield  {author} {\bibinfo {author} {\bibfnamefont {N.}~\bibnamefont
  {Boulanger}},\ }\bibfield  {title} {\bibinfo {title} {{General solutions of
  the Wess-Zumino consistency condition for the Weyl anomalies}},\ }\href
  {https://doi.org/10.1088/1126-6708/2007/07/069} {\bibfield  {journal}
  {\bibinfo  {journal} {JHEP}\ }\textbf {\bibinfo {volume} {07}},\ \bibinfo
  {pages} {069}},\ \Eprint {https://arxiv.org/abs/0704.2472} {arXiv:0704.2472
  [hep-th]} \BibitemShut {NoStop}%
\bibitem [{\citenamefont {Boulanger}(2007{\natexlab{b}})}]{Boulanger:2007ab}%
  \BibitemOpen
  \bibfield  {author} {\bibinfo {author} {\bibfnamefont {N.}~\bibnamefont
  {Boulanger}},\ }\bibfield  {title} {\bibinfo {title} {{Algebraic
  Classification of Weyl Anomalies in Arbitrary Dimensions}},\ }\href
  {https://doi.org/10.1103/PhysRevLett.98.261302} {\bibfield  {journal}
  {\bibinfo  {journal} {Phys. Rev. Lett.}\ }\textbf {\bibinfo {volume} {98}},\
  \bibinfo {pages} {261302} (\bibinfo {year} {2007}{\natexlab{b}})},\ \Eprint
  {https://arxiv.org/abs/0706.0340} {arXiv:0706.0340 [hep-th]} \BibitemShut
  {NoStop}%
\bibitem [{\citenamefont {Duff}(2017)}]{Duff2017}%
  \BibitemOpen
  \bibfield  {author} {\bibinfo {author} {\bibfnamefont {M.~J.}\ \bibnamefont
  {Duff}},\ }\href {{https://pitpas1.phas.ubc.ca/gravity2017/talks/duff.pdf}}
  {\bibinfo {title} {{Five decades of the gravitational Weyl anomaly}}},\
  \bibinfo {howpublished} {Talk given at GRAVITY: Past, Present \& Future,
  University of British Columbia} (\bibinfo {year} {2017})\BibitemShut
  {NoStop}%
\bibitem [{\citenamefont {Osborn}\ and\ \citenamefont
  {Petkou}(1994)}]{Osborn:1993cr}%
  \BibitemOpen
  \bibfield  {author} {\bibinfo {author} {\bibfnamefont {H.}~\bibnamefont
  {Osborn}}\ and\ \bibinfo {author} {\bibfnamefont {A.~C.}\ \bibnamefont
  {Petkou}},\ }\bibfield  {title} {\bibinfo {title} {{Implications of conformal
  invariance in field theories for general dimensions}},\ }\href
  {https://doi.org/10.1006/aphy.1994.1045} {\bibfield  {journal} {\bibinfo
  {journal} {Annals Phys.}\ }\textbf {\bibinfo {volume} {231}},\ \bibinfo
  {pages} {311} (\bibinfo {year} {1994})},\ \Eprint
  {https://arxiv.org/abs/hep-th/9307010} {arXiv:hep-th/9307010} \BibitemShut
  {NoStop}%
\bibitem [{\citenamefont {Beccaria}\ and\ \citenamefont
  {Tseytlin}(2016)}]{Beccaria:2015ypa}%
  \BibitemOpen
  \bibfield  {author} {\bibinfo {author} {\bibfnamefont {M.}~\bibnamefont
  {Beccaria}}\ and\ \bibinfo {author} {\bibfnamefont {A.~A.}\ \bibnamefont
  {Tseytlin}},\ }\bibfield  {title} {\bibinfo {title} {{Conformal anomaly
  c-coefficients of superconformal 6d theories}},\ }\href
  {https://doi.org/10.1007/JHEP01(2016)001} {\bibfield  {journal} {\bibinfo
  {journal} {JHEP}\ }\textbf {\bibinfo {volume} {01}},\ \bibinfo {pages}
  {001}},\ \Eprint {https://arxiv.org/abs/1510.02685} {arXiv:1510.02685
  [hep-th]} \BibitemShut {NoStop}%
\bibitem [{\citenamefont {Henningson}\ and\ \citenamefont
  {Skenderis}(2000)}]{Henningson:1998ey}%
  \BibitemOpen
  \bibfield  {author} {\bibinfo {author} {\bibfnamefont {M.}~\bibnamefont
  {Henningson}}\ and\ \bibinfo {author} {\bibfnamefont {K.}~\bibnamefont
  {Skenderis}},\ }\bibfield  {title} {\bibinfo {title} {{Holography and the
  Weyl anomaly}},\ }\href
  {https://doi.org/10.1002/(SICI)1521-3978(20001)48:1/3<125::AID-PROP125>3.0.CO;2-B}
  {\bibfield  {journal} {\bibinfo  {journal} {Fortsch. Phys.}\ }\textbf
  {\bibinfo {volume} {48}},\ \bibinfo {pages} {125} (\bibinfo {year} {2000})},\
  \Eprint {https://arxiv.org/abs/hep-th/9812032} {arXiv:hep-th/9812032}
  \BibitemShut {NoStop}%
\bibitem [{\citenamefont {Henningson}\ and\ \citenamefont
  {Skenderis}(1998)}]{Henningson:1998gx}%
  \BibitemOpen
  \bibfield  {author} {\bibinfo {author} {\bibfnamefont {M.}~\bibnamefont
  {Henningson}}\ and\ \bibinfo {author} {\bibfnamefont {K.}~\bibnamefont
  {Skenderis}},\ }\bibfield  {title} {\bibinfo {title} {{The Holographic Weyl
  anomaly}},\ }\href {https://doi.org/10.1088/1126-6708/1998/07/023} {\bibfield
   {journal} {\bibinfo  {journal} {JHEP}\ }\textbf {\bibinfo {volume} {07}},\
  \bibinfo {pages} {023}},\ \Eprint {https://arxiv.org/abs/hep-th/9806087}
  {arXiv:hep-th/9806087} \BibitemShut {NoStop}%
\bibitem [{\citenamefont {Gover}\ and\ \citenamefont
  {Peterson}(2003)}]{Gover:2002ay}%
  \BibitemOpen
  \bibfield  {author} {\bibinfo {author} {\bibfnamefont {A.~R.}\ \bibnamefont
  {Gover}}\ and\ \bibinfo {author} {\bibfnamefont {L.~J.}\ \bibnamefont
  {Peterson}},\ }\bibfield  {title} {\bibinfo {title} {{Conformally invariant
  powers of the Laplacian, Q-curvature, and tractor calculus}},\ }\href
  {https://doi.org/10.1007/s00220-002-0790-4} {\bibfield  {journal} {\bibinfo
  {journal} {Commun. Math. Phys.}\ }\textbf {\bibinfo {volume} {235}},\
  \bibinfo {pages} {339} (\bibinfo {year} {2003})},\ \Eprint
  {https://arxiv.org/abs/math-ph/0201030} {arXiv:math-ph/0201030} \BibitemShut
  {NoStop}%
\bibitem [{\citenamefont {Jia}\ and\ \citenamefont
  {Karydas}(2021)}]{Jia:2021hgy}%
  \BibitemOpen
  \bibfield  {author} {\bibinfo {author} {\bibfnamefont {W.}~\bibnamefont
  {Jia}}\ and\ \bibinfo {author} {\bibfnamefont {M.}~\bibnamefont {Karydas}},\
  }\bibfield  {title} {\bibinfo {title} {{Obstruction tensors in Weyl geometry
  and holographic Weyl anomaly}},\ }\href
  {https://doi.org/10.1103/PhysRevD.104.126031} {\bibfield  {journal} {\bibinfo
   {journal} {Phys. Rev. D}\ }\textbf {\bibinfo {volume} {104}},\ \bibinfo
  {pages} {126031} (\bibinfo {year} {2021})},\ \Eprint
  {https://arxiv.org/abs/2109.14014} {arXiv:2109.14014 [hep-th]} \BibitemShut
  {NoStop}%
\bibitem [{\citenamefont {Boulanger}\ and\ \citenamefont
  {Rovere}(2026)}]{Boulanger:2025oli}%
  \BibitemOpen
  \bibfield  {author} {\bibinfo {author} {\bibfnamefont {N.}~\bibnamefont
  {Boulanger}}\ and\ \bibinfo {author} {\bibfnamefont {D.}~\bibnamefont
  {Rovere}},\ }\bibfield  {title} {\bibinfo {title} {{8D conformal gravity with
  Einstein sector, and its relation to the Q-curvature}},\ }\href
  {https://doi.org/10.1007/JHEP02(2026)101} {\bibfield  {journal} {\bibinfo
  {journal} {JHEP}\ }\textbf {\bibinfo {volume} {02}},\ \bibinfo {pages}
  {101}},\ \Eprint {https://arxiv.org/abs/2511.01368} {arXiv:2511.01368
  [hep-th]} \BibitemShut {NoStop}%
\bibitem [{\citenamefont {Graham}\ and\ \citenamefont
  {Zworski}(2003)}]{Graham2003}%
  \BibitemOpen
  \bibfield  {author} {\bibinfo {author} {\bibfnamefont {C.~R.}\ \bibnamefont
  {Graham}}\ and\ \bibinfo {author} {\bibfnamefont {M.}~\bibnamefont
  {Zworski}},\ }\bibfield  {title} {\bibinfo {title} {{Scattering matrix in
  conformal geometry}},\ }\href {https://doi.org/10.1007/s00222-002-0268-1}
  {\bibfield  {journal} {\bibinfo  {journal} {Invent. Math.}\ }\textbf
  {\bibinfo {volume} {152}},\ \bibinfo {pages} {89} (\bibinfo {year} {2003})},\
  \Eprint {https://arxiv.org/abs/math/0109089} {arXiv:math/0109089 [math.DG]}
  \BibitemShut {NoStop}%
\bibitem [{\citenamefont {Chen}\ and\ \citenamefont {Lu}(2025)}]{Chen:2024kuw}%
  \BibitemOpen
  \bibfield  {author} {\bibinfo {author} {\bibfnamefont {F.-Y.}\ \bibnamefont
  {Chen}}\ and\ \bibinfo {author} {\bibfnamefont {H.}~\bibnamefont {Lu}},\
  }\bibfield  {title} {\bibinfo {title} {{Holographic Weyl anomaly in 8D from
  general higher curvature gravity}},\ }\href
  {https://doi.org/10.1103/PhysRevD.111.086032} {\bibfield  {journal} {\bibinfo
   {journal} {Phys. Rev. D}\ }\textbf {\bibinfo {volume} {111}},\ \bibinfo
  {pages} {086032} (\bibinfo {year} {2025})},\ \Eprint
  {https://arxiv.org/abs/2410.16097} {arXiv:2410.16097 [hep-th]} \BibitemShut
  {NoStop}%
\bibitem [{\citenamefont {Boulanger}\ and\ \citenamefont
  {Erdmenger}(2004)}]{Boulanger:2004zf}%
  \BibitemOpen
  \bibfield  {author} {\bibinfo {author} {\bibfnamefont {N.}~\bibnamefont
  {Boulanger}}\ and\ \bibinfo {author} {\bibfnamefont {J.}~\bibnamefont
  {Erdmenger}},\ }\bibfield  {title} {\bibinfo {title} {{A Classification of
  local Weyl invariants in $D$ = 8}},\ }\href
  {https://doi.org/10.1088/0264-9381/21/18/003} {\bibfield  {journal} {\bibinfo
   {journal} {Class. Quant. Grav.}\ }\textbf {\bibinfo {volume} {21}},\
  \bibinfo {pages} {4305} (\bibinfo {year} {2004})},\ \bibinfo {note}
  {[Erratum: Class.Quant.Grav. 39, 039501 (2022)]},\ \Eprint
  {https://arxiv.org/abs/hep-th/0405228} {arXiv:hep-th/0405228} \BibitemShut
  {NoStop}%
\bibitem [{\citenamefont {Case}\ \emph {et~al.}(2026)\citenamefont {Case},
  \citenamefont {Khaitan}, \citenamefont {Lin}, \citenamefont {Tyrrell},\ and\
  \citenamefont {Yuan}}]{Case:2024oih}%
  \BibitemOpen
  \bibfield  {author} {\bibinfo {author} {\bibfnamefont {J.~S.}\ \bibnamefont
  {Case}}, \bibinfo {author} {\bibfnamefont {A.}~\bibnamefont {Khaitan}},
  \bibinfo {author} {\bibfnamefont {Y.-J.}\ \bibnamefont {Lin}}, \bibinfo
  {author} {\bibfnamefont {A.~J.}\ \bibnamefont {Tyrrell}},\ and\ \bibinfo
  {author} {\bibfnamefont {W.}~\bibnamefont {Yuan}},\ }\bibfield  {title}
  {\bibinfo {title} {{Computing renormalized curvature integrals on
  Poincar{\'e}-Einstein manifolds}},\ }\href
  {https://doi.org/10.1016/j.aim.2026.110991} {\bibfield  {journal} {\bibinfo
  {journal} {Adv. Math.}\ }\textbf {\bibinfo {volume} {496}},\ \bibinfo {pages}
  {110991} (\bibinfo {year} {2026})},\ \Eprint
  {https://arxiv.org/abs/2404.11319} {arXiv:2404.11319 [math.DG]} \BibitemShut
  {NoStop}%
\bibitem [{\citenamefont {Liu}\ and\ \citenamefont
  {Tseytlin}(1998)}]{Liu:1998bu}%
  \BibitemOpen
  \bibfield  {author} {\bibinfo {author} {\bibfnamefont {H.}~\bibnamefont
  {Liu}}\ and\ \bibinfo {author} {\bibfnamefont {A.~A.}\ \bibnamefont
  {Tseytlin}},\ }\bibfield  {title} {\bibinfo {title} {{D = 4 superYang-Mills,
  D = 5 gauged supergravity, and D = 4 conformal supergravity}},\ }\href
  {https://doi.org/10.1016/S0550-3213(98)00443-X} {\bibfield  {journal}
  {\bibinfo  {journal} {Nucl. Phys. B}\ }\textbf {\bibinfo {volume} {533}},\
  \bibinfo {pages} {88} (\bibinfo {year} {1998})},\ \Eprint
  {https://arxiv.org/abs/hep-th/9804083} {arXiv:hep-th/9804083} \BibitemShut
  {NoStop}%
\bibitem [{\citenamefont {Bastianelli}\ \emph {et~al.}(2000)\citenamefont
  {Bastianelli}, \citenamefont {Frolov},\ and\ \citenamefont
  {Tseytlin}}]{Bastianelli:2000hi}%
  \BibitemOpen
  \bibfield  {author} {\bibinfo {author} {\bibfnamefont {F.}~\bibnamefont
  {Bastianelli}}, \bibinfo {author} {\bibfnamefont {S.}~\bibnamefont
  {Frolov}},\ and\ \bibinfo {author} {\bibfnamefont {A.~A.}\ \bibnamefont
  {Tseytlin}},\ }\bibfield  {title} {\bibinfo {title} {{Conformal anomaly of
  (2,0) tensor multiplet in six-dimensions and AdS / CFT correspondence}},\
  }\href {https://doi.org/10.1088/1126-6708/2000/02/013} {\bibfield  {journal}
  {\bibinfo  {journal} {JHEP}\ }\textbf {\bibinfo {volume} {02}},\ \bibinfo
  {pages} {013}},\ \Eprint {https://arxiv.org/abs/hep-th/0001041}
  {arXiv:hep-th/0001041} \BibitemShut {NoStop}%
\bibitem [{\citenamefont {Erdmenger}\ and\ \citenamefont
  {Osborn}(1997)}]{Erdmenger:1996yc}%
  \BibitemOpen
  \bibfield  {author} {\bibinfo {author} {\bibfnamefont {J.}~\bibnamefont
  {Erdmenger}}\ and\ \bibinfo {author} {\bibfnamefont {H.}~\bibnamefont
  {Osborn}},\ }\bibfield  {title} {\bibinfo {title} {{Conserved currents and
  the energy momentum tensor in conformally invariant theories for general
  dimensions}},\ }\href {https://doi.org/10.1016/S0550-3213(96)00545-7}
  {\bibfield  {journal} {\bibinfo  {journal} {Nucl. Phys. B}\ }\textbf
  {\bibinfo {volume} {483}},\ \bibinfo {pages} {431} (\bibinfo {year}
  {1997})},\ \Eprint {https://arxiv.org/abs/hep-th/9605009}
  {arXiv:hep-th/9605009} \BibitemShut {NoStop}%
\bibitem [{\citenamefont {Hofman}\ and\ \citenamefont
  {Maldacena}(2008)}]{Hofman:2008ar}%
  \BibitemOpen
  \bibfield  {author} {\bibinfo {author} {\bibfnamefont {D.~M.}\ \bibnamefont
  {Hofman}}\ and\ \bibinfo {author} {\bibfnamefont {J.}~\bibnamefont
  {Maldacena}},\ }\bibfield  {title} {\bibinfo {title} {{Conformal collider
  physics: Energy and charge correlations}},\ }\href
  {https://doi.org/10.1088/1126-6708/2008/05/012} {\bibfield  {journal}
  {\bibinfo  {journal} {JHEP}\ }\textbf {\bibinfo {volume} {05}},\ \bibinfo
  {pages} {012}},\ \Eprint {https://arxiv.org/abs/0803.1467} {arXiv:0803.1467
  [hep-th]} \BibitemShut {NoStop}%
\bibitem [{\citenamefont {Buchel}\ \emph {et~al.}(2010)\citenamefont {Buchel},
  \citenamefont {Escobedo}, \citenamefont {Myers}, \citenamefont {Paulos},
  \citenamefont {Sinha},\ and\ \citenamefont {Smolkin}}]{Buchel:2009sk}%
  \BibitemOpen
  \bibfield  {author} {\bibinfo {author} {\bibfnamefont {A.}~\bibnamefont
  {Buchel}}, \bibinfo {author} {\bibfnamefont {J.}~\bibnamefont {Escobedo}},
  \bibinfo {author} {\bibfnamefont {R.~C.}\ \bibnamefont {Myers}}, \bibinfo
  {author} {\bibfnamefont {M.~F.}\ \bibnamefont {Paulos}}, \bibinfo {author}
  {\bibfnamefont {A.}~\bibnamefont {Sinha}},\ and\ \bibinfo {author}
  {\bibfnamefont {M.}~\bibnamefont {Smolkin}},\ }\bibfield  {title} {\bibinfo
  {title} {{Holographic GB gravity in arbitrary dimensions}},\ }\href
  {https://doi.org/10.1007/JHEP03(2010)111} {\bibfield  {journal} {\bibinfo
  {journal} {JHEP}\ }\textbf {\bibinfo {volume} {03}},\ \bibinfo {pages}
  {111}},\ \Eprint {https://arxiv.org/abs/0911.4257} {arXiv:0911.4257 [hep-th]}
  \BibitemShut {NoStop}%
\bibitem [{\citenamefont {Chen}\ and\ \citenamefont {Lu}(2024)}]{Chen:2023bgj}%
  \BibitemOpen
  \bibfield  {author} {\bibinfo {author} {\bibfnamefont {F.-Y.}\ \bibnamefont
  {Chen}}\ and\ \bibinfo {author} {\bibfnamefont {H.}~\bibnamefont {Lu}},\
  }\bibfield  {title} {\bibinfo {title} {{Holographic three-point functions
  from higher curvature gravities in arbitrary dimensions}},\ }\href
  {https://doi.org/10.1103/PhysRevD.109.064070} {\bibfield  {journal} {\bibinfo
   {journal} {Phys. Rev. D}\ }\textbf {\bibinfo {volume} {109}},\ \bibinfo
  {pages} {064070} (\bibinfo {year} {2024})},\ \Eprint
  {https://arxiv.org/abs/2311.05307} {arXiv:2311.05307 [hep-th]} \BibitemShut
  {NoStop}%
\bibitem [{\citenamefont {Gibbons}(1999)}]{Gibbons:1998zr}%
  \BibitemOpen
  \bibfield  {author} {\bibinfo {author} {\bibfnamefont {G.~W.}\ \bibnamefont
  {Gibbons}},\ }\bibfield  {title} {\bibinfo {title} {{Some comments on
  gravitational entropy and the inverse mean curvature flow}},\ }\href
  {https://doi.org/10.1088/0264-9381/16/6/302} {\bibfield  {journal} {\bibinfo
  {journal} {Class. Quant. Grav.}\ }\textbf {\bibinfo {volume} {16}},\ \bibinfo
  {pages} {1677} (\bibinfo {year} {1999})},\ \Eprint
  {https://arxiv.org/abs/hep-th/9809167} {arXiv:hep-th/9809167} \BibitemShut
  {NoStop}%
\bibitem [{\citenamefont {Susskind}\ and\ \citenamefont
  {Witten}(1998)}]{Susskind:1998dq}%
  \BibitemOpen
  \bibfield  {author} {\bibinfo {author} {\bibfnamefont {L.}~\bibnamefont
  {Susskind}}\ and\ \bibinfo {author} {\bibfnamefont {E.}~\bibnamefont
  {Witten}},\ }\bibfield  {title} {\bibinfo {title} {{The Holographic bound in
  anti-de Sitter space}},\ }\href@noop {} {\  (\bibinfo {year} {1998})},\
  \Eprint {https://arxiv.org/abs/hep-th/9805114} {arXiv:hep-th/9805114}
  \BibitemShut {NoStop}%
\bibitem [{\citenamefont {Besse}(1987)}]{Besse:1987pua}%
  \BibitemOpen
  \bibfield  {author} {\bibinfo {author} {\bibfnamefont {A.~L.}\ \bibnamefont
  {Besse}},\ }\href@noop {} {\emph {\bibinfo {title} {{Einstein Manifolds}}}}\
  (\bibinfo  {publisher} {Springer-Verlag},\ \bibinfo {address} {Berlin,
  Heidelberg, New York},\ \bibinfo {year} {1987})\BibitemShut {NoStop}%
\bibitem [{\citenamefont {Bzowski}\ \emph {et~al.}(2018)\citenamefont
  {Bzowski}, \citenamefont {McFadden},\ and\ \citenamefont
  {Skenderis}}]{Bzowski:2017poo}%
  \BibitemOpen
  \bibfield  {author} {\bibinfo {author} {\bibfnamefont {A.}~\bibnamefont
  {Bzowski}}, \bibinfo {author} {\bibfnamefont {P.}~\bibnamefont {McFadden}},\
  and\ \bibinfo {author} {\bibfnamefont {K.}~\bibnamefont {Skenderis}},\
  }\bibfield  {title} {\bibinfo {title} {{Renormalised 3-point functions of
  stress tensors and conserved currents in CFT}},\ }\href
  {https://doi.org/10.1007/JHEP11(2018)153} {\bibfield  {journal} {\bibinfo
  {journal} {JHEP}\ }\textbf {\bibinfo {volume} {11}},\ \bibinfo {pages}
  {153}},\ \Eprint {https://arxiv.org/abs/1711.09105} {arXiv:1711.09105
  [hep-th]} \BibitemShut {NoStop}%
\bibitem [{\citenamefont {Dunne}\ and\ \citenamefont
  {Rius}(1992)}]{Dunne:1992ws}%
  \BibitemOpen
  \bibfield  {author} {\bibinfo {author} {\bibfnamefont {G.~V.}\ \bibnamefont
  {Dunne}}\ and\ \bibinfo {author} {\bibfnamefont {N.}~\bibnamefont {Rius}},\
  }\bibfield  {title} {\bibinfo {title} {{A Comment on the relationship between
  differential and dimensional renormalization}},\ }\href
  {https://doi.org/10.1016/0370-2693(92)90897-D} {\bibfield  {journal}
  {\bibinfo  {journal} {Phys. Lett. B}\ }\textbf {\bibinfo {volume} {293}},\
  \bibinfo {pages} {367} (\bibinfo {year} {1992})},\ \Eprint
  {https://arxiv.org/abs/hep-th/9206038} {arXiv:hep-th/9206038} \BibitemShut
  {NoStop}%
\bibitem [{\citenamefont {Gel'fand}\ and\ \citenamefont
  {Shilov}(1964)}]{Gelfand1964}%
  \BibitemOpen
  \bibfield  {author} {\bibinfo {author} {\bibfnamefont {I.~M.}\ \bibnamefont
  {Gel'fand}}\ and\ \bibinfo {author} {\bibfnamefont {G.~E.}\ \bibnamefont
  {Shilov}},\ }\href@noop {} {\emph {\bibinfo {title} {{Generalized Functions:
  Properties and Operations, Vol. 1}}}}\ (\bibinfo  {publisher} {Academic
  Press},\ \bibinfo {address} {New York},\ \bibinfo {year} {1964})\ \bibinfo
  {note} {translated from the Russian edition of 1958}\BibitemShut {NoStop}%
\bibitem [{\citenamefont {Duff}(2020)}]{Duff:2020dqb}%
  \BibitemOpen
  \bibfield  {author} {\bibinfo {author} {\bibfnamefont {M.~J.}\ \bibnamefont
  {Duff}},\ }\bibfield  {title} {\bibinfo {title} {{Weyl, Pontryagin, Euler,
  Eguchi and Freund}},\ }\href {https://doi.org/10.1088/1751-8121/ab956d}
  {\bibfield  {journal} {\bibinfo  {journal} {J. Phys. A}\ }\textbf {\bibinfo
  {volume} {53}},\ \bibinfo {pages} {301001} (\bibinfo {year} {2020})},\
  \Eprint {https://arxiv.org/abs/2003.02688} {arXiv:2003.02688 [hep-th]}
  \BibitemShut {NoStop}%
\bibitem [{\citenamefont {Diaz}(2008)}]{Diaz:2008hy}%
  \BibitemOpen
  \bibfield  {author} {\bibinfo {author} {\bibfnamefont {D.~E.}\ \bibnamefont
  {Diaz}},\ }\bibfield  {title} {\bibinfo {title} {{Polyakov formulas for GJMS
  operators from AdS/CFT}},\ }\href
  {https://doi.org/10.1088/1126-6708/2008/07/103} {\bibfield  {journal}
  {\bibinfo  {journal} {JHEP}\ }\textbf {\bibinfo {volume} {07}},\ \bibinfo
  {pages} {103}},\ \Eprint {https://arxiv.org/abs/0803.0571} {arXiv:0803.0571
  [hep-th]} \BibitemShut {NoStop}%
\bibitem [{\citenamefont {Schwimmer}\ and\ \citenamefont
  {Theisen}(2019)}]{Schwimmer:2019efk}%
  \BibitemOpen
  \bibfield  {author} {\bibinfo {author} {\bibfnamefont {A.}~\bibnamefont
  {Schwimmer}}\ and\ \bibinfo {author} {\bibfnamefont {S.}~\bibnamefont
  {Theisen}},\ }\bibfield  {title} {\bibinfo {title} {{Osborn Equation and
  Irrelevant Operators}},\ }\href {https://doi.org/10.1088/1742-5468/ab3284}
  {\bibfield  {journal} {\bibinfo  {journal} {J. Stat. Mech.}\ }\textbf
  {\bibinfo {volume} {1908}},\ \bibinfo {pages} {084011} (\bibinfo {year}
  {2019})},\ \Eprint {https://arxiv.org/abs/1902.04473} {arXiv:1902.04473
  [hep-th]} \BibitemShut {NoStop}%
\bibitem [{\citenamefont {Sen}\ and\ \citenamefont
  {Sinha}(2014)}]{Sen:2014nfa}%
  \BibitemOpen
  \bibfield  {author} {\bibinfo {author} {\bibfnamefont {K.}~\bibnamefont
  {Sen}}\ and\ \bibinfo {author} {\bibfnamefont {A.}~\bibnamefont {Sinha}},\
  }\bibfield  {title} {\bibinfo {title} {{Holographic stress tensor at finite
  coupling}},\ }\href {https://doi.org/10.1007/JHEP07(2014)098} {\bibfield
  {journal} {\bibinfo  {journal} {JHEP}\ }\textbf {\bibinfo {volume} {07}},\
  \bibinfo {pages} {098}},\ \Eprint {https://arxiv.org/abs/1405.7862}
  {arXiv:1405.7862 [hep-th]} \BibitemShut {NoStop}%
\bibitem [{\citenamefont {Osborn}\ and\ \citenamefont
  {Stergiou}(2015)}]{Osborn:2015rna}%
  \BibitemOpen
  \bibfield  {author} {\bibinfo {author} {\bibfnamefont {H.}~\bibnamefont
  {Osborn}}\ and\ \bibinfo {author} {\bibfnamefont {A.}~\bibnamefont
  {Stergiou}},\ }\bibfield  {title} {\bibinfo {title} {{Structures on the
  Conformal Manifold in Six Dimensional Theories}},\ }\href
  {https://doi.org/10.1007/JHEP04(2015)157} {\bibfield  {journal} {\bibinfo
  {journal} {JHEP}\ }\textbf {\bibinfo {volume} {04}},\ \bibinfo {pages}
  {157}},\ \Eprint {https://arxiv.org/abs/1501.01308} {arXiv:1501.01308
  [hep-th]} \BibitemShut {NoStop}%
\bibitem [{\citenamefont {L{\"u}}\ \emph {et~al.}(2013)\citenamefont {L{\"u}},
  \citenamefont {Pang},\ and\ \citenamefont {Pope}}]{Lu:2013hx}%
  \BibitemOpen
  \bibfield  {author} {\bibinfo {author} {\bibfnamefont {H.}~\bibnamefont
  {L{\"u}}}, \bibinfo {author} {\bibfnamefont {Y.}~\bibnamefont {Pang}},\ and\
  \bibinfo {author} {\bibfnamefont {C.~N.}\ \bibnamefont {Pope}},\ }\bibfield
  {title} {\bibinfo {title} {{Black Holes in Six-dimensional Conformal
  Gravity}},\ }\href {https://doi.org/10.1103/PhysRevD.87.104013} {\bibfield
  {journal} {\bibinfo  {journal} {Phys. Rev. D}\ }\textbf {\bibinfo {volume}
  {87}},\ \bibinfo {pages} {104013} (\bibinfo {year} {2013})},\ \Eprint
  {https://arxiv.org/abs/1301.7083} {arXiv:1301.7083 [hep-th]} \BibitemShut
  {NoStop}%
\bibitem [{\citenamefont {Erdmenger}\ and\ \citenamefont
  {Osborn}(1998)}]{Erdmenger:1997wy}%
  \BibitemOpen
  \bibfield  {author} {\bibinfo {author} {\bibfnamefont {J.}~\bibnamefont
  {Erdmenger}}\ and\ \bibinfo {author} {\bibfnamefont {H.}~\bibnamefont
  {Osborn}},\ }\bibfield  {title} {\bibinfo {title} {{Conformally covariant
  differential operators: Symmetric tensor fields}},\ }\href
  {https://doi.org/10.1088/0264-9381/15/2/003} {\bibfield  {journal} {\bibinfo
  {journal} {Class. Quant. Grav.}\ }\textbf {\bibinfo {volume} {15}},\ \bibinfo
  {pages} {273} (\bibinfo {year} {1998})},\ \Eprint
  {https://arxiv.org/abs/gr-qc/9708040} {arXiv:gr-qc/9708040} \BibitemShut
  {NoStop}%
\bibitem [{\citenamefont {Graham}\ and\ \citenamefont
  {Hirachi}(2005)}]{Graham2005}%
  \BibitemOpen
  \bibfield  {author} {\bibinfo {author} {\bibfnamefont {C.~R.}\ \bibnamefont
  {Graham}}\ and\ \bibinfo {author} {\bibfnamefont {K.}~\bibnamefont
  {Hirachi}},\ }\bibinfo {title} {{The ambient obstruction tensor and
  Q-curvature}},\ in\ \href {https://doi.org/10.4171/013-1/3} {\emph {\bibinfo
  {booktitle} {AdS/CFT Correspondence: Einstein Metrics and Their Conformal
  Boundaries}}}\ (\bibinfo  {publisher} {EMS Press},\ \bibinfo {year} {2005})\
  pp.\ \bibinfo {pages} {59--71},\ \Eprint {https://arxiv.org/abs/math/0405068}
  {arXiv:math/0405068 [math.DG]} \BibitemShut {NoStop}%
\bibitem [{\citenamefont {Matsumoto}(2013)}]{Matsumoto2013}%
  \BibitemOpen
  \bibfield  {author} {\bibinfo {author} {\bibfnamefont {Y.}~\bibnamefont
  {Matsumoto}},\ }\bibfield  {title} {\bibinfo {title} {{A GJMS construction
  for 2-tensors and the second variation of the total Q-curvature}},\ }\href
  {https://doi.org/10.2140/pjm.2013.262.437} {\bibfield  {journal} {\bibinfo
  {journal} {Pac. J. Math.}\ }\textbf {\bibinfo {volume} {262}},\ \bibinfo
  {pages} {437} (\bibinfo {year} {2013})},\ \Eprint
  {https://arxiv.org/abs/1202.3227} {arXiv:1202.3227 [math.DG]} \BibitemShut
  {NoStop}%
\bibitem [{\citenamefont {Beccaria}\ and\ \citenamefont
  {Tseytlin}(2015{\natexlab{a}})}]{Beccaria:2015vaa}%
  \BibitemOpen
  \bibfield  {author} {\bibinfo {author} {\bibfnamefont {M.}~\bibnamefont
  {Beccaria}}\ and\ \bibinfo {author} {\bibfnamefont {A.~A.}\ \bibnamefont
  {Tseytlin}},\ }\bibfield  {title} {\bibinfo {title} {{On higher spin
  partition functions}},\ }\href
  {https://doi.org/10.1088/1751-8113/48/27/275401} {\bibfield  {journal}
  {\bibinfo  {journal} {J. Phys. A}\ }\textbf {\bibinfo {volume} {48}},\
  \bibinfo {pages} {275401} (\bibinfo {year} {2015}{\natexlab{a}})},\ \Eprint
  {https://arxiv.org/abs/1503.08143} {arXiv:1503.08143 [hep-th]} \BibitemShut
  {NoStop}%
\bibitem [{\citenamefont {Beccaria}\ and\ \citenamefont
  {Tseytlin}(2015{\natexlab{b}})}]{Beccaria:2015uta}%
  \BibitemOpen
  \bibfield  {author} {\bibinfo {author} {\bibfnamefont {M.}~\bibnamefont
  {Beccaria}}\ and\ \bibinfo {author} {\bibfnamefont {A.~A.}\ \bibnamefont
  {Tseytlin}},\ }\bibfield  {title} {\bibinfo {title} {{Conformal a-anomaly of
  some non-unitary 6d superconformal theories}},\ }\href
  {https://doi.org/10.1007/JHEP09(2015)017} {\bibfield  {journal} {\bibinfo
  {journal} {JHEP}\ }\textbf {\bibinfo {volume} {09}},\ \bibinfo {pages}
  {017}},\ \Eprint {https://arxiv.org/abs/1506.08727} {arXiv:1506.08727
  [hep-th]} \BibitemShut {NoStop}%
\bibitem [{\citenamefont {Osborn}\ and\ \citenamefont
  {Stergiou}(2016)}]{Osborn:2016bev}%
  \BibitemOpen
  \bibfield  {author} {\bibinfo {author} {\bibfnamefont {H.}~\bibnamefont
  {Osborn}}\ and\ \bibinfo {author} {\bibfnamefont {A.}~\bibnamefont
  {Stergiou}},\ }\bibfield  {title} {\bibinfo {title} {{C$_{T}$ for non-unitary
  CFTs in higher dimensions}},\ }\href
  {https://doi.org/10.1007/JHEP06(2016)079} {\bibfield  {journal} {\bibinfo
  {journal} {JHEP}\ }\textbf {\bibinfo {volume} {06}},\ \bibinfo {pages}
  {079}},\ \Eprint {https://arxiv.org/abs/1603.07307} {arXiv:1603.07307
  [hep-th]} \BibitemShut {NoStop}%
\bibitem [{\citenamefont {Guerrieri}\ \emph {et~al.}(2016)\citenamefont
  {Guerrieri}, \citenamefont {Petkou},\ and\ \citenamefont
  {Wen}}]{Guerrieri:2016whh}%
  \BibitemOpen
  \bibfield  {author} {\bibinfo {author} {\bibfnamefont {A.}~\bibnamefont
  {Guerrieri}}, \bibinfo {author} {\bibfnamefont {A.~C.}\ \bibnamefont
  {Petkou}},\ and\ \bibinfo {author} {\bibfnamefont {C.}~\bibnamefont {Wen}},\
  }\bibfield  {title} {\bibinfo {title} {{The free $\sigma$CFTs}},\ }\href
  {https://doi.org/10.1007/JHEP09(2016)019} {\bibfield  {journal} {\bibinfo
  {journal} {JHEP}\ }\textbf {\bibinfo {volume} {09}},\ \bibinfo {pages}
  {019}},\ \Eprint {https://arxiv.org/abs/1604.07310} {arXiv:1604.07310
  [hep-th]} \BibitemShut {NoStop}%
\bibitem [{\citenamefont {Avramidi}(1991)}]{Avramidi:1990je}%
  \BibitemOpen
  \bibfield  {author} {\bibinfo {author} {\bibfnamefont {I.~G.}\ \bibnamefont
  {Avramidi}},\ }\bibfield  {title} {\bibinfo {title} {{The Covariant Technique
  for Calculation of One Loop Effective Action}},\ }\href
  {https://doi.org/10.1016/0550-3213(91)90492-G} {\bibfield  {journal}
  {\bibinfo  {journal} {Nucl. Phys. B}\ }\textbf {\bibinfo {volume} {355}},\
  \bibinfo {pages} {712} (\bibinfo {year} {1991})},\ \bibinfo {note} {[Erratum:
  Nucl.Phys.B 509, 557--558 (1998)]}\BibitemShut {NoStop}%
\bibitem [{\citenamefont {Aros}\ \emph {et~al.}(2026)\citenamefont {Aros},
  \citenamefont {Bugini}, \citenamefont {D{\'i}az},\ and\ \citenamefont
  {N{\'u}{\~n}ez-Barra}}]{Aros2026}%
  \BibitemOpen
  \bibfield  {author} {\bibinfo {author} {\bibfnamefont {R.}~\bibnamefont
  {Aros}}, \bibinfo {author} {\bibfnamefont {F.}~\bibnamefont {Bugini}},
  \bibinfo {author} {\bibfnamefont {D.~E.}\ \bibnamefont {D{\'i}az}},\ and\
  \bibinfo {author} {\bibfnamefont {C.}~\bibnamefont {N{\'u}{\~n}ez-Barra}},\
  }\href@noop {} {\bibinfo {title} {{Simple recipe for 8d holographic Weyl
  anomaly}}},\ \bibinfo {howpublished} {In preparation} (\bibinfo {year}
  {2026})\BibitemShut {NoStop}%
\bibitem [{\citenamefont {Bugini}\ and\ \citenamefont
  {D{\'\i}az}(2019)}]{Bugini:2018def}%
  \BibitemOpen
  \bibfield  {author} {\bibinfo {author} {\bibfnamefont {F.}~\bibnamefont
  {Bugini}}\ and\ \bibinfo {author} {\bibfnamefont {D.~E.}\ \bibnamefont
  {D{\'\i}az}},\ }\bibfield  {title} {\bibinfo {title} {{Holographic Weyl
  anomaly for GJMS operators: one Laplacian to rule them all}},\ }\href
  {https://doi.org/10.1007/JHEP02(2019)188} {\bibfield  {journal} {\bibinfo
  {journal} {JHEP}\ }\textbf {\bibinfo {volume} {02}},\ \bibinfo {pages}
  {188}},\ \Eprint {https://arxiv.org/abs/1811.10380} {arXiv:1811.10380
  [hep-th]} \BibitemShut {NoStop}%
\bibitem [{\citenamefont {Hawking}(1977)}]{Hawking:1976ja}%
  \BibitemOpen
  \bibfield  {author} {\bibinfo {author} {\bibfnamefont {S.~W.}\ \bibnamefont
  {Hawking}},\ }\bibfield  {title} {\bibinfo {title} {{Zeta Function
  Regularization of Path Integrals in Curved Space-Time}},\ }\href
  {https://doi.org/10.1007/BF01626516} {\bibfield  {journal} {\bibinfo
  {journal} {Commun. Math. Phys.}\ }\textbf {\bibinfo {volume} {55}},\ \bibinfo
  {pages} {133} (\bibinfo {year} {1977})}\BibitemShut {NoStop}%
\bibitem [{\citenamefont {Duff}(1994)}]{Duff:1993wm}%
  \BibitemOpen
  \bibfield  {author} {\bibinfo {author} {\bibfnamefont {M.~J.}\ \bibnamefont
  {Duff}},\ }\bibfield  {title} {\bibinfo {title} {{Twenty years of the Weyl
  anomaly}},\ }\href {https://doi.org/10.1088/0264-9381/11/6/004} {\bibfield
  {journal} {\bibinfo  {journal} {Class. Quant. Grav.}\ }\textbf {\bibinfo
  {volume} {11}},\ \bibinfo {pages} {1387} (\bibinfo {year} {1994})},\ \Eprint
  {https://arxiv.org/abs/hep-th/9308075} {arXiv:hep-th/9308075} \BibitemShut
  {NoStop}%
\bibitem [{\citenamefont {Bugini}\ and\ \citenamefont
  {Diaz}(2017)}]{Bugini:2016nvn}%
  \BibitemOpen
  \bibfield  {author} {\bibinfo {author} {\bibfnamefont {F.}~\bibnamefont
  {Bugini}}\ and\ \bibinfo {author} {\bibfnamefont {D.~E.}\ \bibnamefont
  {Diaz}},\ }\bibfield  {title} {\bibinfo {title} {{Simple recipe for
  holographic Weyl anomaly}},\ }\href {https://doi.org/10.1007/JHEP04(2017)122}
  {\bibfield  {journal} {\bibinfo  {journal} {JHEP}\ }\textbf {\bibinfo
  {volume} {04}},\ \bibinfo {pages} {122}},\ \Eprint
  {https://arxiv.org/abs/1612.00351} {arXiv:1612.00351 [hep-th]} \BibitemShut
  {NoStop}%
\bibitem [{\citenamefont {Imbimbo}\ \emph {et~al.}(2000)\citenamefont
  {Imbimbo}, \citenamefont {Schwimmer}, \citenamefont {Theisen},\ and\
  \citenamefont {Yankielowicz}}]{Imbimbo:1999bj}%
  \BibitemOpen
  \bibfield  {author} {\bibinfo {author} {\bibfnamefont {C.}~\bibnamefont
  {Imbimbo}}, \bibinfo {author} {\bibfnamefont {A.}~\bibnamefont {Schwimmer}},
  \bibinfo {author} {\bibfnamefont {S.}~\bibnamefont {Theisen}},\ and\ \bibinfo
  {author} {\bibfnamefont {S.}~\bibnamefont {Yankielowicz}},\ }\bibfield
  {title} {\bibinfo {title} {{Diffeomorphisms and holographic anomalies}},\
  }\href {https://doi.org/10.1088/0264-9381/17/5/322} {\bibfield  {journal}
  {\bibinfo  {journal} {Class. Quant. Grav.}\ }\textbf {\bibinfo {volume}
  {17}},\ \bibinfo {pages} {1129} (\bibinfo {year} {2000})},\ \Eprint
  {https://arxiv.org/abs/hep-th/9910267} {arXiv:hep-th/9910267} \BibitemShut
  {NoStop}%
\bibitem [{\citenamefont {Bueno}\ \emph {et~al.}(2019)\citenamefont {Bueno},
  \citenamefont {Cano}, \citenamefont {Hennigar},\ and\ \citenamefont
  {Mann}}]{Bueno:2018yzo}%
  \BibitemOpen
  \bibfield  {author} {\bibinfo {author} {\bibfnamefont {P.}~\bibnamefont
  {Bueno}}, \bibinfo {author} {\bibfnamefont {P.~A.}\ \bibnamefont {Cano}},
  \bibinfo {author} {\bibfnamefont {R.~A.}\ \bibnamefont {Hennigar}},\ and\
  \bibinfo {author} {\bibfnamefont {R.~B.}\ \bibnamefont {Mann}},\ }\bibfield
  {title} {\bibinfo {title} {{Universality of Squashed-Sphere Partition
  Functions}},\ }\href {https://doi.org/10.1103/PhysRevLett.122.071602}
  {\bibfield  {journal} {\bibinfo  {journal} {Phys. Rev. Lett.}\ }\textbf
  {\bibinfo {volume} {122}},\ \bibinfo {pages} {071602} (\bibinfo {year}
  {2019})},\ \Eprint {https://arxiv.org/abs/1808.02052} {arXiv:1808.02052
  [hep-th]} \BibitemShut {NoStop}%
\bibitem [{\citenamefont {Li}\ \emph {et~al.}(2018)\citenamefont {Li},
  \citenamefont {L{\"u}},\ and\ \citenamefont {Mai}}]{Li:2018drw}%
  \BibitemOpen
  \bibfield  {author} {\bibinfo {author} {\bibfnamefont {Y.-Z.}\ \bibnamefont
  {Li}}, \bibinfo {author} {\bibfnamefont {H.}~\bibnamefont {L{\"u}}},\ and\
  \bibinfo {author} {\bibfnamefont {Z.-F.}\ \bibnamefont {Mai}},\ }\bibfield
  {title} {\bibinfo {title} {{Universal Structure of Covariant Holographic
  Two-Point Functions In Massless Higher-Order Gravities}},\ }\href
  {https://doi.org/10.1007/JHEP10(2018)063} {\bibfield  {journal} {\bibinfo
  {journal} {JHEP}\ }\textbf {\bibinfo {volume} {10}},\ \bibinfo {pages}
  {063}},\ \Eprint {https://arxiv.org/abs/1808.00494} {arXiv:1808.00494
  [hep-th]} \BibitemShut {NoStop}%
\bibitem [{\citenamefont {de~Boer}\ \emph {et~al.}(2010)\citenamefont
  {de~Boer}, \citenamefont {Kulaxizi},\ and\ \citenamefont
  {Parnachev}}]{deBoer:2009pn}%
  \BibitemOpen
  \bibfield  {author} {\bibinfo {author} {\bibfnamefont {J.}~\bibnamefont
  {de~Boer}}, \bibinfo {author} {\bibfnamefont {M.}~\bibnamefont {Kulaxizi}},\
  and\ \bibinfo {author} {\bibfnamefont {A.}~\bibnamefont {Parnachev}},\
  }\bibfield  {title} {\bibinfo {title} {{AdS(7)/CFT(6), Gauss-Bonnet Gravity,
  and Viscosity Bound}},\ }\href {https://doi.org/10.1007/JHEP03(2010)087}
  {\bibfield  {journal} {\bibinfo  {journal} {JHEP}\ }\textbf {\bibinfo
  {volume} {03}},\ \bibinfo {pages} {087}},\ \Eprint
  {https://arxiv.org/abs/0910.5347} {arXiv:0910.5347 [hep-th]} \BibitemShut
  {NoStop}%
\bibitem [{\citenamefont {Perlmutter}(2014)}]{Perlmutter:2013gua}%
  \BibitemOpen
  \bibfield  {author} {\bibinfo {author} {\bibfnamefont {E.}~\bibnamefont
  {Perlmutter}},\ }\bibfield  {title} {\bibinfo {title} {{A universal feature
  of CFT R{\'e}nyi entropy}},\ }\href {https://doi.org/10.1007/JHEP03(2014)117}
  {\bibfield  {journal} {\bibinfo  {journal} {JHEP}\ }\textbf {\bibinfo
  {volume} {03}},\ \bibinfo {pages} {117}},\ \Eprint
  {https://arxiv.org/abs/1308.1083} {arXiv:1308.1083 [hep-th]} \BibitemShut
  {NoStop}%
\bibitem [{\citenamefont {Beccaria}\ and\ \citenamefont
  {Tseytlin}(2017)}]{Beccaria:2017lcz}%
  \BibitemOpen
  \bibfield  {author} {\bibinfo {author} {\bibfnamefont {M.}~\bibnamefont
  {Beccaria}}\ and\ \bibinfo {author} {\bibfnamefont {A.~A.}\ \bibnamefont
  {Tseytlin}},\ }\bibfield  {title} {\bibinfo {title} {{C$_{T}$ for conformal
  higher spin fields from partition function on conically deformed sphere}},\
  }\href {https://doi.org/10.1007/JHEP09(2017)123} {\bibfield  {journal}
  {\bibinfo  {journal} {JHEP}\ }\textbf {\bibinfo {volume} {09}},\ \bibinfo
  {pages} {123}},\ \Eprint {https://arxiv.org/abs/1707.02456} {arXiv:1707.02456
  [hep-th]} \BibitemShut {NoStop}%
\bibitem [{\citenamefont {Anastasiou}\ \emph {et~al.}(2025)\citenamefont
  {Anastasiou}, \citenamefont {Araya}, \citenamefont {Das},\ and\ \citenamefont
  {Moreno}}]{Anastasiou:2025rvz}%
  \BibitemOpen
  \bibfield  {author} {\bibinfo {author} {\bibfnamefont {G.}~\bibnamefont
  {Anastasiou}}, \bibinfo {author} {\bibfnamefont {I.~J.}\ \bibnamefont
  {Araya}}, \bibinfo {author} {\bibfnamefont {A.}~\bibnamefont {Das}},\ and\
  \bibinfo {author} {\bibfnamefont {J.}~\bibnamefont {Moreno}},\ }\bibfield
  {title} {\bibinfo {title} {{Universality of pseudoentropy for deformed
  spheres in dS/CFT}},\ }\href@noop {} {\  (\bibinfo {year} {2025})},\ \Eprint
  {https://arxiv.org/abs/2512.02164} {arXiv:2512.02164 [hep-th]} \BibitemShut
  {NoStop}%
\end{thebibliography}

%

\end{document}